\documentclass[11pt,a4paper]{article}
\usepackage{tikz}
\usepackage{authblk}
\usepackage{epsfig}
\usepackage{subfigure}
\usepackage{amsmath}
\usepackage{amssymb}
\usepackage{ulem}
\usepackage{mathtools}
\usepackage{graphicx}
\usepackage{MnSymbol}
\graphicspath{{plots/}}
\usepackage[hidelinks]{hyperref}
\usepackage{csquotes}
\usepackage{multirow}
\usepackage[left=.45in,right=.45in,top=.6in,bottom=.6in,headheight=14.5pt]{geometry}
\usepackage{multirow}
\usepackage{xcolor}
\usepackage{pdflscape}
\usepackage{orcidlink}
\usepackage[title]{appendix}
\usepackage[left=.45in,right=.45in,top=.6in,bottom=.6in,headheight=14.5pt]{geometry}
\usepackage{fmtcount} 
\usepackage{array,multirow}
\newcolumntype{C}[1]{>{\centering\arraybackslash}m{#1}}
\usepackage{epsfig}
\usepackage{float}
\usepackage{geometry}
\usepackage{amsmath}
\usepackage{cite}
\usepackage{ctable}
\newgeometry{vmargin={10mm,20mm},hmargin={19mm,19mm}}
\title{On Lepton Flavor Violation and Dark Matter in Scotogenic model with Trimaximal Mixing}

\author[1]{Tapender\thanks{tapenderphy@gmail.com}}
\author[1] {Surender Verma\thanks{s\_7verma@hpcu.ac.in}}
\author[2]{Sanjeev Kumar\thanks{ skverma@physics.du.ac.in}}

\affil[1]{Department of Physics and Astronomical Science, Central University of Himachal Pradesh, Dharamshala-176215, INDIA}
\affil[2]{Department of Physics and Astrophysics, University of Delhi, Delhi-110007, INDIA}

\date{}

\begin{document}

\maketitle
\begin{abstract}
\noindent We examine the Scotogenic model employing the TM$_2$ mixing matrix, $U_{\text{TM}_2}$, for neutrinos and parameterize the Yukawa coupling matrix $y$ based on the diagonalization condition for the neutrino mass matrix, $m_{\nu}$. Our investigation centers on analyzing the relic density of cold dark matter ($\Omega h^2$) and possible lepton flavor violation (LFV) in the model. In particular, we study coannihilation dynamics and LFV, in the model, considering various coannihilation scenarios including non-zero mass splitting between lightest sterile neutrinos. While analyzing, we have taken into consideration respective experimental constraints on $\Omega h^2$ and LFV alongside neutrino oscillation data. Our study reveals that in both normal and inverted hierarchy of neutrino masses, splitting between masses of $N_1$ and $N_2$ can be up to $\approx 15\%$ for the model to be in consonance with the above constraints. In the second part, we have extended the analysis incorporating extended magic symmetry in $m_\nu$ enabling us to completely determine Yukawa coupling matrix ($y$).  We observe a notable exclusion of the effective Majorana mass $|m_{ee}|$ parameter space by cosmological bound on sum of neutrino masses, particularly in the normal hierarchy while inverted hierarchy scenario is excluded due to constraints coming from extended magic symmetry. These findings shed light on the interplay among the Scotogenic model, TM$_2$ mixing, and extended magic symmetry, offering insights into the permitted parameter space and hierarchy exclusion.
\end{abstract}
 
\section{Introduction}
The experimental observations during the past two decades have affirmed our belief in the Standard Model (SM) of particle physics as the effective low energy theory explaining particle interactions at the fundamental level. Despite our present knowledge that neutrino has tiny mass and they mix while propagating in space, their theoretical origin is still not understood. Further, complete pattern of the neutrino mixing matrix is, also, unknown. Alongside observables in the neutrino sector, discerning cosmological dark matter (DM) and possible lepton flavor violation (LFV) emanating from underlying theory constitute formidable challenge in particle physics. 

\noindent In this context, various extensions  of the SM have been proposed to account, simultaneously, for non-zero neutrino mass and DM. Out of these extensions, the \textit{Scotogenic model} stands as a theoretical framework in particle physics, providing a unified solution to two fundamental mysteries: the origin of neutrino masses and the existence of dark matter\cite{Ma:2006km}. It was developed to address the puzzle of small neutrino masses and to offer insights into dark matter. In the Scotogenic model, neutrino masses arise at the 1-loop level, driven by quantum corrections involving new particles beyond the Standard Model. These additional particles, typically scalar or fermionic fields, interact with known particles in the Standard Model. The lightest of the $Z_2$ odd parity state is the possible DM particle in the model. In these models, the dark matter candidate originates from the same sector responsible for generating neutrino masses, establishing a natural connection between these phenomena\cite{Ma:2008cu,Farzan:2009ji,Farzan:2010mr,Hehn:2012kz,Brdar:2013iea,Chao:2012sz,vonderPahlen:2016cbw,Rocha-Moran:2016enp,Ibarra:2016dlb,Ferreira:2016sbb,Lu:2016dbc,Mahanta:2019gfe,Borah:2018rca,Vicente:2014wga,Ghosh:2022fws,Chun:2023vbh,Fortes:2017ndr,Borah:2021khc,Singh:2023eye,Cacciapaglia:2020psm,Rosenlyst:2021tdr,Rosenlyst:2022jxj,Karan:2023adm}. 

\noindent Meanwhile, the observation of a non-zero value for $\theta_{13}$ in various neutrino oscillation experiments has led to a reassessment of the Tribimaximal (TBM) mixing pattern\cite{Harrison:1999cf,Harrison:2002er,Xing:2002sw,Harrison:2002kp,Harrison:2003aw,He:2003rm,Li:2004dn} in light of experimental data 
 \cite{Xing:2011at,Zhou:2011nu,Araki:2011wn,Haba:2011nv,Chao:2011sp,Zhang:2011aw,Rodejohann:2011uz,Marzocca:2011dh,Antusch:2011ic,Dev:2011bd,Ge:2011qn,Ge:2011ih,Ludl:2011vv,Joshipura:2011rr,Morisi:2011pm,BhupalDev:2011gi,deAdelhartToorop:2011nfg,Adulpravitchai:2011rq,Cao:2011cp,Araki:2011qy,Rashed:2011xe,Rashed:2011zs,Aranda:2011rt,Meloni:2011ac,King:2011ab,Kashav:2023tmz,Zhao:2017yvw,Channey:2021pjf}.  An alternative approach has emerged, preserving one column of the TBM mixing matrix while adjusting the remaining two to meet unitarity constraints. This adaptation results in three distinct \textit{trimaximal} (TM) mixing patterns: TM$_1$, TM$_2$, and TM$_3$ \cite{Haba:2006dz,He:2006qd,Grimus:2008tt,Ishimori:2010fs,Shimizu:2011xg,He:2011gb,deMedeirosVarzielas:2012cet,Zhao:2020cjm,King:2019vhv,Novichkov:2018yse,Rodejohann:2017lre,Luhn:2013lkn,King:2011zj,Kumar:2010qz,Dev:2022krz,Grimus:2009xw,Albright:2008rp,Ding:2020vud,Zhao:2021efc,Rodejohann:2012cf,Zhang:2024rwv,Ganguly:2023jml,Chen:2023hmn}, where the unchanged columns correspond to the first, second, and third positions within the TBM matrix, respectively. While the TM$_3$ configuration predicts $\theta_{13}$ to be zero, rendering it incompatible with observed data, both the TM$_1$ and TM$_2$ schemes have garnered significant attention for their ability to elucidate lepton mixing behaviors. The TM$_2$ scheme, in particular, aligns well with current neutrino oscillation data. Recent research has explored neutrino mass matrices incorporating TM$_1$ and TM$_2$ mixing frameworks with other phenomenological patterns of neutrino mass matrix, contributing to a deeper understanding of neutrino mixing phenomena\cite{Gautam:2016qyw,Kumar:2017hjn,Ding:2013bpa,Loualidi:2021qoq,Gautam:2018izb,Kumar:2023iaj,Mazumder:2023ate,Singh:2022nmk,Channey:2018cfj,Yang:2021xob}.

\noindent One of the key element in Scotogenesis is the structure of the Yukawa matrix ($y$) which  is responsible for generating non-zero neutrino mass, possible LFV and contributes to relic density of the DM in the model. Alternative approaches have been employed in the literature to constrain its structure, based on (i) assuming specific parametrization of the Yukawa coupling matrix ($y$)\cite{Casas:2001sr} (ii) assuming appropriate patterns of the neutrino mixing matrix consistent with current neutrino oscillation data\cite{Suematsu:2009ww, Suematsu:2010gv, Ho:2013hia, Ho:2013spa, Singirala:2016kam,Kitabayashi:2018bye,Schmidt:2012yg, Kitabayashi:2019uzg,Ankush:2021opd,Kitabayashi:2017sjz,Ankush:2023pax}, to name a few. It is pertinent to note that these approaches may have different predictions for observables like relic density of DM and LFV. Particularly, assuming an appropriate structure of neutrino mixing matrix may, further, constrain the allowed parameter space of the model satisfying observational constraints on neutrino oscillation parameters\cite{deSalas:2020pgw}, LFV bounds\cite{MEG:2016leq,SINDRUMII:2006dvw,BaBar:2009hkt,Hayasaka:2010np,SINDRUM:1987nra,SINDRUMII:1993gxf} and relic density of DM\cite{Planck:2018vyg}.

Dark matter has, also, been  investigated  earlier  for example, the works reported in Refs. \cite{Ma:2008cu,Chao:2012sz,Ibarra:2016dlb,Mahanta:2019gfe,Borah:2021khc,Vicente:2014wga,Ghosh:2022fws,Chun:2023vbh,Karan:2023adm,Ho:2013hia,Ho:2013spa,Singirala:2016kam,Suematsu:2009ww, Suematsu:2010gv,Schmidt:2012yg,Kitabayashi:2018bye,Kitabayashi:2019uzg,Ankush:2021opd,Kitabayashi:2017sjz,Ankush:2023pax} deals with fermionic dark matter while in Refs. \cite{Farzan:2009ji,Farzan:2010mr,Brdar:2013iea,vonderPahlen:2016cbw,Lu:2016dbc,Borah:2018rca,Singh:2023eye} scalar dark matter has been investigated in scotogenic framework. The fermionic dark matter in scotogenic scenario has, also, been discussed in Ref. \cite{Suematsu:2009ww} emphasizing the importance of coannihilations in obtaining successful relic density of dark matter while simultaneously satisfying experimental constraint on LFV in  $\mu\rightarrow e \gamma$ process. It is difficult to remove tension between successful relic density of DM and branching ratio of LFV processes in a single framework as both observables depend on neutrino Yukawa couplings in a non-trivial manner. So, constraining structure of Yukawa coupling may play an important role especially when there are  large number of free parameters in the model. In order to reduce parameter in the model some of the studies \cite{ Ho:2013spa, Singirala:2016kam,Suematsu:2009ww, Suematsu:2010gv,Kitabayashi:2018bye} have assumed real neutrino Yukawa couplings. Thus, these studies are generally silent on possible CP-violation, Majorana phases and related observables like effective Majorana mass $|m_{ee}|$. Motivated by this, we, in this study, have 
investigated the relationship between neutrino physics and dark matter by employing TM$_2$ neutrino mixing matrix and considering different coannihilation scenarios. The TM$_2$ mixing pattern result in magic neutrino mass matrix in which the sum of elements of each row/column is equal\cite{Lam:2006wy,Harrison:2004he,Friedberg:2006it,Verma:2019uiu,Hyodo:2022xyn,Hyodo:2021kqf,Hyodo:2020ysq,Minamizawa:2022fch}. We utilize the diagonalization condition of the flavor neutrino mass matrix to derive constraints on the elements of complex Yukawa coupling matrix ($y$). By integrating experimental constraints from both the neutrino sector and dark matter studies, we constrain the parameter space. We have, also, investigated the prediction of the model for possible  LFV in $\mu\rightarrow e\gamma$ process on which we have the stringent experimental upper bound $\mathcal{O}(10^{-13})$\cite{MEG:2016leq}. Additionally, we have obtained predictions for CP-violation, Majorana phases, and effective Majorana mass $|m_{ee}|$. Furthermore, we incorporate the extended magic symmetry, proposed recently in Ref. \cite{Singh:2022nmk}, to further constrain the allowed parameter space. Our analysis indicates that the inverted hierarchy of neutrinos is not allowed under the constraint of the extended magic symmetry.

The paper is organized as follows: in Section \ref{s2}, we have discussed the basic ingredients of the Scotogenic framework. In Section \ref{s3}, we present analytical details of LFV and DM studies in the Scotogenic setup. In Section \ref{s4}, we have discussed our formalism resulting in specific pattern of Yukawa coupling matrix ($y$) emanating from trimaximal mixing pattern (TM$_2$) of the neutrino mixing matrix. The numerical analysis and discussion is presented in Section \ref{s5}. In Section \ref{s6}, we extend our analysis to the extended magic symmetry case for both normal and inverted hierarchies of neutrino masses. Finally, we present conclusions of the present work in Section \ref{s7}.

\section{The Scotogenic Model}{\label{s2}}

Among the myriad extensions proposed for the Standard Model (SM), the Scotogenic model \cite{Ma:2006km} stands out for its unique capability to concurrently account for both neutrino mass and existence  of dark matter. In its conventional form, the Scotogenic model encompasses the complete set of SM fields alongside three additional singlet Majorana fields, denoted as $N_k$ $(k=1,\, 2,\, 3)$, and an inert SU(2)$_L$ scalar doublet, represented as $\zeta=$($\zeta^+$, $\zeta^0$), where $\zeta^+$ is charged component and $\zeta^0$ is the neutral component. From a symmetry perspective, in addition to the SM symmetry group (SU(3)$_C \times$ SU(2)$_L \times$U(1)$_Y$), the model introduces a new $Z_2$ symmetry. Under this exact $Z_2$ symmetry, all newly introduced fields possess odd parity, while all SM fields exhibit even parity viz.,
\[ N_k\xrightarrow{} -N_k,\,\,\, \zeta\xrightarrow{}-\zeta,\,\,\, \Phi\xrightarrow{}\Phi,\,\,\, \Psi_{\text{SM}}\xrightarrow{}\Psi_{\text{SM}},\]
where $\Psi_{\text{SM}}$ denotes all the SM fermions and $\Phi$ denotes SM Higgs doublet.

For the Scotogenic model, the relevant symmetry group comprises SU(2)$_L \times$U(1)$_Y \times$ $Z_2$. In the context of the lepton sector, the pertinent part of the Yukawa Lagrangian can be expressed as follows
\begin{equation}\label{lag1}
-\mathcal{L}_Y= \Gamma_{\alpha \beta} \overline{L}_{\alpha} \Phi l_{R \beta}+y_{\alpha k} \overline{L}_{\alpha} \tilde{\zeta} N_k+\frac{1}{2} M_{k}\overline{N^c_k} N_k+\text{H.c.}\,\,,
\end{equation}
where   $\alpha$, $\beta=$ $e$, $\mu$, $\tau$ represent charged leptons,  $M_k$ denote the mass of three Majorana singlet fermions $N_k$, $(k = 1,\, 2,\, 3)$. The symbols $L_{\alpha}$ and $l_{R \beta}$ denote left-handed  lepton doublets and right-handed lepton singlets, respectively, while $\Gamma_{\alpha \beta}$ and $y_{\alpha k}$ denote Yukawa couplings for charged leptons and neutrinos, respectively. Also, $\tilde{\zeta}=i \sigma_2 \zeta^*$ where $\sigma_2$ is  Pauli spin matrix and $N^c_k=C\overline{N_k}^T$, where $C$ is charge conjugation matrix.

The scalar potential under the exact $Z_2$ symmetry can be written as
\begin{equation}\label{pot1}
\begin{split}
V(\Phi, \zeta) = & \,\, \mu^2_{1}\Phi^{\dagger}\Phi+m^2_{\zeta}\zeta^{\dagger}\zeta+\frac{1}{2}\lambda_1(\Phi^{\dagger}\Phi)^2+\frac{1}{2}\lambda_2(\zeta^{\dagger}\zeta)^2  \\
& +\lambda_3(\Phi^{\dagger}\Phi)(\zeta^{\dagger}\zeta)+\lambda_4(\Phi^{\dagger}\zeta) (\zeta^{\dagger}\Phi) \\
& +\frac{1}{2}\lambda_5\big[(\Phi^{\dagger}\zeta)^2 +\text{H.c.}\big] \,,
\end{split}
\end{equation}
where $\mu_1$, $\mu_2$, and $\lambda_1$ through $\lambda_5$ are  real parameters.\\
As indicated by the imposed symmetry under $Z_2$, the presence of a term involving the Dirac Yukawa couplings of the $N_k$ with the SM Higgs doublet $\Phi$ is precluded. Additionally, since $\zeta$ is classified as an inert doublet, it does not acquire a vacuum expectation value (\textit{vev}), thereby obstructing the generation of light neutrino mass at tree level through  type-I seesaw mechanism. Consequently, neutrinos persist without acquiring mass at tree level. However, they can acquire mass through one-loop interactions, where the quartic term in the scalar potential assumes significance. 
 
After electroweak symmetry breaking, the light neutrino mass generated at one-loop level can be expressed as 
 \begin{equation}\label{mv1}
     (m_{\nu})_{\alpha \beta}=\sum_{k=1}^3 y_{\alpha k} y_{\beta k} \Lambda_k=(y\Lambda y^T)_{\alpha \beta} \,, (  \alpha, \beta= e, \mu, \tau).
 \end{equation}
In Eqn. (\ref{mv1}), $y$ denotes the Yukawa coupling matrix which, in general, can be written as
\begin{equation}\label{yu1}
y =\begin{pmatrix}
y_{e 1}& y_{e 2} & y_{e 3}\\
y_{\mu 1} &  y_{\mu 2}& y_{\mu 3} \\
y_{ \tau 1} & y_{ \tau 2}  & y_{ \tau 3} \\
\end{pmatrix},
\end{equation}
and $\Lambda= diag(\Lambda_1, \Lambda_2, \Lambda_3)$, is the diagonal matrix with elements  
\begin{equation}\label{lam1}
     \Lambda_{k}= \frac{\lambda_5 v^2}{16 \pi^2}\frac{1}{m_0}\frac{r_k}{1-r_k^2}\left(1-\frac{r_k^2}{1-r_k^2} \ln\frac{1}{r_k^2}\right),
 \end{equation}
where $r_k\equiv \frac{M_k}{m_{0}}$, $v$ is the $\textit{vev}$ of the SM Higgs doublet  and $m_0^2=\frac{1}{2}(m_R^2+m_I^2)$ where $m_R$ and $m_I$ denote masses of $\sqrt{2} \Re[\zeta^0]$ and $\sqrt{2} \Im [\zeta^0]$,
respectively. It is to be noted that $\Lambda_k>0$ for $r_k\neq1$.
The radiatively generated neutrino masses will be small when we have $|\lambda_5|<<1$. 

\section{Lepton Flavor Violation and Coannihilation of Dark Matter}{\label{s3}}

In the Scotogenic model, the exact $Z_2$ symmetry ensures that the lightest $Z_2$-odd particle remains stable, rendering it a viable candidate for dark matter. Thus, two scenarios emerge: the dark matter particle can either be a fermion or a neutral scalar. In the fermionic case, the lightest $Z_2$-odd particle would be the dark matter candidate, denoted as $N_1$. Alternatively, if the dark matter candidate is a neutral scalar, it would correspond to the component $\zeta^0$ of the inert doublet $\zeta$. 

For the purpose of this study, we assume fermionic dark matter, implying that $N_1$ serves as the dark matter particle. This assumption sets the stage for investigating the properties and interactions of $N_1$ within the Scotogenic model framework.


Lepton flavor violation (LFV) is a phenomenon actively sought after in numerous experiments, with the most popular searches being for the radiative process is  $l_\alpha \rightarrow l_\beta \gamma$  involving rare decay of a charged lepton into another  charged lepton and a photon. In the Scotogenic model, this process is induced  at the one-loop level. 
The branching ratio of $l_\alpha \rightarrow l_\beta \gamma$ is given by\cite{Toma:2013zsa}
\begin{equation}
	\text{Br}(l_\alpha \rightarrow l_\beta \gamma)=\frac{3 (4 \pi)^3 \alpha_{em}}{4 G_F^2}\left|A_D\right|^2 	\text{Br}(l_\alpha \rightarrow l_\beta \nu_\alpha \overline{\nu}_\beta)\,,
\end{equation}
where $\alpha_{em}$ denotes the fine-structure constant, $G_F$ denotes the Fermi coupling constant, and $A_D$  is dipole form factor given by
\begin{equation}
    A_D=\sum_{k=1}^3 \frac{y_{\alpha k} y_{\beta k}^*}{2 (4 \pi)^2}\frac{1}{m_0^2}F(r_k)\,,
\end{equation}
where $\lambda_4<<\lambda_3$,  thus,  $\zeta^+$ and $\zeta^0$ becomes degenerate having mass $m_0$. The loop function $F(r_k)$ is defined by
\begin{equation}
	F(r_k)=\frac{1-6r_k^2+3r_k^4+2r_k^6-6r_k^4 \ln r_k^2}{6(1-r_k^2)^4}.
\end{equation}
Among various radiative LFV transitions, $\mu\rightarrow e \gamma$ has most stringent constraint (Br$(\mu\rightarrow e \gamma)\leq 4.2 \times 10^{-13}$ \cite{MEG:2016leq}) so we aim to investigate both the relic density of cold dark matter (CDM) and LFV using bounds coming from $\mu\rightarrow e \gamma$ process, but tension arises when considering a singlet fermion as a candidate for cold dark matter \cite{Kubo:2006yx}. This tension arises because the same Yukawa couplings must satisfy experimental bounds for both processes, yet each process requires a different strength of the Yukawa couplings. This discrepancy poses a challenge in reconciling the requirements of both the relic density and the LFV process within the framework of a singlet fermion dark matter model.
 
To simultaneously study CDM and LFV within this model, considering the references \cite{Suematsu:2009ww,Suematsu:2010gv}, it's essential to recognize that when coannihilation effects are considered, the predicted CDM density and the branching ratio of the LFV process $\mu \rightarrow e \gamma$ can both align with observations. This alignment holds within the simplest Scotogenic model. Coannihilation  happens if other particles heavier than dark matter particle are close in mass to the dark matter particle \cite{Griest:1990kh}. It  lead to various scenarios. The cases that we are interested in are as follows:
	\begin{enumerate}
		\item $M_1\approx M_2<m_0$: In this case, $N_1$ and $N_2$ are nearly degenerate i.e., splitting $\delta M=(M_2-M_1)/M_1\approx 0$, and the other particles are heavier. So, here we can see the effect of coannihilation between $N_1$ and $N_2$, in this process only Yukawa couplings are important. 
		\item $M_1\approx M_2\lesssim m_0$: Here, $N_1$ and $N_2$ are nearly degenerate in mass, similar to case (1), but $\zeta$ is also close in mass to $N_1$. So, here we study the $N_1$, $N_2$, and $\zeta$ coannihilations, in which gauge interactions are expected to play an important role. 
		\item $M_1< M_2< m_0$: In this case, we assume non-zero splitting between $N_1$ and $N_2$ with $\delta M$ in the range  of $(10\%-20\%)$, while the rest remains the same as in  case $(2)$. Here,  we  investigate the impact of splitting on coannihilation between $N_1$, $N_2$, and $\zeta$.
	\end{enumerate}
 In all of the above cases, $M_3$ is the largest mass.
In this work, we have used SARAH-4.15.2 \cite{Staub:2008uz,Staub:2013tta,Staub:2015kfa,Porod:2014xia}  to generate modules for  SPheno-4.0.5 \cite{Porod:2003um,Porod:2011nf} and micrOMEGAs-5.3.41 \cite{Belanger:2001fz,Belanger:2020gnr,Belanger:2021smw,Alguero:2022inz,Belanger:2014vza} to numerically calculate LFV observables and CDM relic density, respectively, which takes account of all relevant processes.

\section{Formalism}{\label{s4}}

In our study, we recognize that the elements of the Yukawa coupling matrix are pivotal for determining both the branching ratio of LFV processes and the relic density of cold dark matter (CDM). Two common approaches for determining the Yukawa coupling matrix are available: 

1. \textbf{Casas-Ibarra (CI) Parametrization}: This method, initially proposed in \cite{Casas:2001sr} and adapted to the Scotogenic model \cite{Toma:2013zsa}, involves parametrizing the Yukawa coupling matrix using CI parametrization. This approach offers a systematic way to link the neutrino sector with LFV processes and CDM relic density calculations.

2. \textbf{Constraints from Diagonalization Condition of Neutrino Mass Matrix}: Alternatively, one can use assumptions regarding the neutrino mixing matrix and apply the diagonalization condition of the flavor neutrino mass matrix to derive constraints on the Yukawa coupling matrix elements. This approach, extensively utilized in \cite{Suematsu:2009ww, Suematsu:2010gv,Schmidt:2012yg, Ho:2013hia, Ho:2013spa,Kitabayashi:2018bye, Singirala:2016kam}, involves expressing the Yukawa coupling matrix elements in terms of three independent Yukawa couplings. Further assumptions, such as texture zeros in the flavor neutrino mass matrix, can then be made to determine the values of these independent Yukawa couplings \cite{Kitabayashi:2018bye, Kitabayashi:2019uzg,Kitabayashi:2017sjz,Ankush:2023pax,Ankush:2021opd}.

In this paper, we opt for the second method. We assume that mass matrix for charged leptons is diagonal. By leveraging constraints originating from the diagonalization condition of flavor neutrino mass matrix  and assuming certain properties of the neutrino sector, such as texture zeros, we aim to determine the Yukawa coupling matrix elements and subsequently investigate their implications for both LFV processes and the relic density of CDM.

Among the various forms of the  trimaximal mixing, we consider 
\begin{equation}\label{tm21}
U_{\text{TM}_2} =\begin{pmatrix}
\sqrt{\frac{2}{3}}\cos\theta & \frac{1}{\sqrt{3}} & \sqrt{\frac{2}{3}}\sin\theta \\
-\frac{\cos\theta}{\sqrt{6}}+\frac{e^{-i\phi}\sin\theta}{\sqrt{2}} &  \frac{1}{\sqrt{3}} & -\frac{\sin\theta}{\sqrt{6}}-\frac{e^{-i\phi}\cos\theta}{\sqrt{2}} \\
-\frac{\cos\theta}{\sqrt{6}}-\frac{e^{-i\phi}\sin\theta}{\sqrt{2}} &  \frac{1}{\sqrt{3}} & -\frac{\sin\theta}{\sqrt{6}}+\frac{e^{-i\phi}\cos\theta}{\sqrt{2}}\\
\end{pmatrix},
\end{equation}
where $\theta$ and $\phi$ are two free parameters. The neutrino mass  matrix  can be diagonalized by using the transformation
\begin{equation}\label{dia1}
   U^T_{\text{TM}_2}  m_{\nu} U_{\text{TM}_2}= m_{\nu}^{d} ,
 \end{equation}
 
where $m_{\nu}^{d}=diag(m_1 ,m_2 ,m_3 )$ is diagonal neutrino mass matrix with eigenvalues $m_i$ ($i=1,2,3$). Substituting Eqns. (\ref{mv1}) and (\ref{tm21}) in Eqn. (\ref{dia1}), and by requiring the  equality in Eqn. 
 (\ref{dia1}) to hold, we have set of six equations, three for off-diagonal elements and three for  diagonal elements. For simplicity, we have assumed $m^{\nu n}=  U^T_{\text{TM}_2}  m_{\nu} U_{\text{TM}_2}$. The three equations for off-diagonal elements are
 
    \begin{scriptsize}
        \begin{eqnarray}
  m^{\nu n}_{12} &=&\sum_{i=1}^3 \left[\sqrt{2}(2 y_{ei}-y_{\mu i}-y_{\tau i})\cos\theta-\sqrt{6}e^{-i \phi}(y_{\tau i}-y_{\mu i})\sin\theta \right]( y_{ei}+y_{\mu i}+y_{\tau i})\Lambda_i=0\,,\label{mvn121}\\
m^{\nu n}_{13} &=& \sum_{i=1}^3 \left[\sqrt{3}e^{i\phi}(2 y_{ei}-y_{\mu i}-y_{\tau i})(y_{\tau i}-y_{\mu i})\cos2\theta -\frac{3}{2}(y_{\tau i}-y_{\mu i})^2 \sin2\theta  +\frac{1}{2}e^{2i\phi}(2 y_{ei}-y_{\mu i}-y_{\tau i})^2 \sin2\theta \right]\Lambda_i=0\,, \label{mvn131}\\
m^{\nu n}_{23} &=&  \sum_{i=1}^3 \left[\sqrt{6} e^{-i \phi}(y_{\tau i}-y_{\mu i}) \cos\theta+\sqrt{2}(2 y_{ei}-y_{\mu i}-y_{\tau i})\sin\theta\right]( y_{ei}+y_{\mu i}+y_{\tau i})\Lambda_i=0\,,\label{mvn231}
\end{eqnarray}
        \end{scriptsize}
and three equations for diagonal elements are
\begin{scriptsize}
       \begin{eqnarray}\label{nondia}
  m^{\nu n}_{11} &=&\sum_{i=1}^3 \frac{1}{6}\left[(2 y_{ei}-y_{\mu i}-y_{\tau i})^2\cos^2\theta+3e^{-2 i \phi}(y_{\tau i}-y_{\mu i})^2\sin^2\theta -\sqrt{3}e^{- i \phi}(y_{\tau i}-y_{\mu i})(2 y_{ei}-y_{\mu i}-y_{\tau i})\sin2\theta\right]\Lambda_i=m_1 \,,\\
m^{\nu n}_{22} &=& \sum_{i=1}^3 \frac{1}{3} ( y_{ei}+y_{\mu i}+y_{\tau i})^2 \Lambda_i=m_2  \,,\\
m^{\nu n}_{33} &=&  \sum_{i=1}^3 \frac{1}{6}\left[(2 y_{ei}-y_{\mu i}-y_{\tau i})^2\sin^2\theta+3e^{-2 i \phi}(y_{\tau i}-y_{\mu i})^2\cos^2\theta +\sqrt{3}e^{- i \phi}(y_{\tau i}-y_{\mu i})(2 y_{ei}-y_{\mu i}-y_{\tau i})\sin2\theta\right]\Lambda_i=m_3 \,.
\end{eqnarray}
\end{scriptsize}

With nine Yukawa couplings and six equations, we find ourselves with three Yukawa couplings that remain independent. We opt to set them as $y_{e i}=y_i$ for $i=1, 2, 3$ \footnote{This parameterization is, also, utilized in similar studies such as \cite{Singirala:2016kam,Kitabayashi:2018bye, Kitabayashi:2019uzg}}. Multiplying Eqn. (\ref{mvn121}) by $\cos\theta$ and Eqn. (\ref{mvn231}) by $\sin\theta$ and adding both of them, we have
\begin{equation}
    \sum_{i=1}^3(2 y_{e i}-y_{\mu i}-y_{\tau i})(y_{e i}+y_{\mu i}+y_{\tau i})=0.
\end{equation}
Now, we have two solutions: either $(2 y_{e i}-y_{\mu i}-y_{\tau i})=0$ or $(y_{e i}+y_{\mu i}+y_{\tau i})=0$.
We substitute these solutions in Eqn. (\ref{mvn131}) and we obtain following results:
\begin{enumerate}
    \item If  $(2 y_{e i}-y_{\mu i}-y_{\tau i})=0$, then 
    \begin{equation}\label{sol1}
   y_{e i}=y_{\mu i}=y_{\tau i}.
   \end{equation}
    \item If $(y_{e i}+y_{\mu i}+y_{\tau i})=0$, solving this equation with Eqn. (\ref{mvn131}) simultaneously for $y_{\mu i}$ and $y_{\tau i}$, we again have two solutions:
   \begin{enumerate}

   \item  
   \begin{equation}\label{sol2a}
        \begin{aligned}
   \begin{rcases}
      y_{\mu i}=&\frac{1}{2}\left[-1+\sqrt{3}\cot 2\theta(-e^{i\phi}+e^{i \phi} \sec 2\theta )\right]y_{e i}\,,\\
 y_{\tau i}=&\frac{1}{2}\left[-1-\sqrt{3}\cot 2\theta(-e^{i\phi}+e^{i \phi} \sec 2\theta )\right]y_{e i}\,,
 \end{rcases}
\end{aligned}
   \end{equation}
  
   \item \begin{equation}\label{sol2b}
        \begin{aligned}
   \begin{rcases}
      y_{\mu i}=&\frac{1}{2}\left[-1-\sqrt{3}\cot 2\theta(e^{i\phi}+e^{i \phi} \sec 2\theta )\right]y_{e i}\,,\\
 y_{\tau i}=&\frac{1}{2}\left[-1+\sqrt{3}\cot 2\theta(e^{i\phi}+e^{i \phi} \sec 2\theta )\right]y_{e i}\, .
 \end{rcases}
\end{aligned}
   \end{equation}
   \end{enumerate}
\end{enumerate}

It is to be noted that we can independently choose any of these solutions for each value of $i\,(i = 1,\, 2,\, 3)$. As all off-diagonal elements vanish with any of these solution sets, we can now determine the solutions for the diagonal elements that satisfy these conditions. The various choices of solution sets will result in different values for the three diagonal equations. In total, we can derive 27 sets of solutions. Among these, 3 sets yield two zero neutrino masses, 18 sets result in one zero neutrino mass, and the remaining 6 sets produce three non-zero neutrino masses.

Assuming that the mass eigenvalues $m_i$ ($i = 1,\, 2,\, 3$) are all non-zero in our analysis, we have selected one of the 6 sets that have non-zero values for all three neutrino masses. We have found that the allowed parameter space obtained for this solution does not exhibit significant variation compared to other solution  sets with three non-zero neutrino masses. A similar analytical approach can be applied to the other solution sets. We can derive the following solution based on the our choices of solution for each $i$: 

\begin{enumerate}
    \item For $i=2$, we choose  solution shown in Eqn. (\ref{sol1}).
    \item For $i=1$, we choose solution given by Eqn. (\ref{sol2a}) and for $i=3$, we choose solution given by Eqn. (\ref{sol2b}). 
\end{enumerate}

So, the Yukawa coupling matrix parameterized in terms of $y_{e i}=y_i$  ($i=1, 2, 3$) is given by
\begin{equation}\label{yu2}
y =\begin{pmatrix}
y_{1}& y_{2} & y_{3}\\
a_1 y_{1} &  y_{ 2}& a_3 y_{ 3} \\
 a_2 y_{  1} & y_{  2}  & a_4 y_{ 3} \\
\end{pmatrix},
\end{equation}
where 
\begin{eqnarray}
 a_1&= &  \frac{1}{2}\left[-1+\sqrt{3}\cot 2\theta(-e^{i\phi}+e^{i \phi} \sec 2\theta )\right],\label{a1}\\
 a_2&=& \frac{1}{2}\left[-1-\sqrt{3}\cot 2\theta(-e^{i\phi}+e^{i \phi} \sec 2\theta )\right] ,\label{a2}\\
 a_3 &=& \frac{1}{2}\left[-1-\sqrt{3}\cot 2\theta(e^{i\phi}+e^{i \phi} \sec 2\theta )\right],\label{a3} \\
 a_4 &=& \frac{1}{2}\left[-1+\sqrt{3}\cot 2\theta(e^{i\phi}+e^{i \phi} \sec 2\theta )\right],\label{a4}
\end{eqnarray}
and neutrino mass eigenvalues are given by
\begin{equation}\label{m1m2m3_1}
    m_1  =c_1 y_1^2 \Lambda_1,\,\,\ m_2 =c_2 y_2^2 \Lambda_2, \,\,\ m_3 =c_3 y_3^2 \Lambda_3,
\end{equation}
where $c_1=\frac{3}{2}\sec^2\theta, c_2=3, c_3=\frac{3}{2}\csc^2\theta$ and $\Lambda_1\approx \Lambda_2.$ As Yukawa couplings are in general complex, we can write $y^2_1=|y^2_1| e^{i2\alpha_1}$, $y^2_2=|y^2_2| e^{i2\alpha_2}$ and $y^2_3=|y^2_3| e^{i2\alpha_3}$ which implies $y_i=\sqrt{|y^2_i|}e^{i\alpha_i}$, where $\alpha_i\,\,(i=1, 2, 3)$ are three Majorana phases. So, we have\footnote{Note that $c_k \, (k=1,2,3)$ is always real and positive and $\Lambda_k>0$  for $r_k\neq1$ is also real.}
\begin{equation}\label{m1m2m3_2}
    m_1  =|c_1 y_1^2 \Lambda_1|e^{i2\alpha_1},\,\,\ m_2 =|c_2 y_2^2 \Lambda_2|e^{i2\alpha_2}, \,\,\ m_3 =|c_3 y_3^2 \Lambda_3|e^{i2\alpha_3}.
\end{equation}
So, we get real and positive masses as
\begin{equation}\label{m1m2m3_3}
    |m_1|  =|c_1 y_1^2 \Lambda_1|,\,\,\ |m_2| =|c_2 y_2^2 \Lambda_2|, \,\,\ |m_3| =|c_3 y_3^2 \Lambda_3|.
\end{equation}
Further these masses need to satisfy mass-squared differences given by
\begin{eqnarray}
    \Delta m^2_{21}&=&|m_2|^2-|m_1|^2, \\
    |\Delta m^2_{31}|&=&||m_3|^2-|m_1|^2|,
\end{eqnarray}
where $\Delta m^2_{31}$ is positive for normal hierarchy (NH)  and negative for inverted hierarchy (IH) of neutrinos. We can write three mass eigenvalues in terms of lightest neutrino mass. For NH, we have
\begin{eqnarray}
    |m_2|&=&\sqrt{\Delta m^2_{21}+|m_1|^2}, \\
    |m_3|&=&\sqrt{\Delta m^2_{31}+|m_1|^2},
\end{eqnarray}
and we can write the $y_2$ and $y_3$ in terms of $y_1$ as
\begin{eqnarray}
    |y_2^2|&=&\frac{1}{|c_2  \Lambda_1|}\sqrt{\Delta m^2_{21}+ |c_1  y_1^2 \Lambda_1 |^2}, \nonumber\\
    y_2&=&\sqrt{\frac{1}{|c_2  \Lambda_1|}\sqrt{\Delta m^2_{21}+ |c_1  y_1^2 \Lambda_1 |^2}}e^{ i \alpha_2},\label{y21}
\end{eqnarray}
and
\begin{eqnarray}
	|y_3^2|&=&\frac{1}{|c_3 \Lambda_3|}\sqrt{\Delta m^2_{31}+ |c_1  y_1^2 \Lambda_1 |^2},\nonumber\\
	y_3&=&\sqrt{\frac{1}{|c_3 \Lambda_3|}\sqrt{\Delta m^2_{31}+ |c_1  y_1^2 \Lambda_1 |^2}}e^{ i\alpha_3}. \label{y31}
\end{eqnarray}
Here, we left with only one independent Yukawa coupling $y_1$ for the NH case. For the IH case, we have
\begin{eqnarray}
    |m_1|&=&\sqrt{\Delta m^2_{31}+|m_3|^2},\\
    |m_2|&=&\sqrt{\Delta m^2_{21}+\Delta m^2_{31}+|m_3|^2}.
\end{eqnarray}
We can write the $y_1$ and $y_2$ in terms of $y_3$ as
\begin{eqnarray}
    |y_1^2|&=&\frac{1}{|c_1  \Lambda_1|}\sqrt{\Delta m^2_{31}+ |c_3  y_3^2 \Lambda_3 |^2},\nonumber\\
    y_1&=&\sqrt{\frac{1}{|c_1  \Lambda_1|}\sqrt{\Delta m^2_{31}+ |c_3  y_3^2 \Lambda_3 |^2}}e^{ i \alpha_1},\label{y13}
     \end{eqnarray}
and
\begin{eqnarray}
	|y_2^2|&=&\frac{1}{|c_2 \Lambda_1|}\sqrt{\Delta m^2_{21}+\Delta m^2_{31}+ |c_3  y_3^2 \Lambda_3 |^2},\nonumber\\
	y_2&=&\sqrt{\frac{1}{|c_2 \Lambda_1|}\sqrt{\Delta m^2_{21}+\Delta m^2_{31}+ |c_3  y_3^2 \Lambda_3 |^2}}e^{ i \alpha_2}.\label{y23}
\end{eqnarray}
So, for the IH case, we  are left with only one independent Yukawa coupling $y_3$. Using Eqn. (\ref{yu2}) in Eqn. (\ref{mv1}), the resulting neutrino mass matrix $m_{\nu}$ can be written as
\begin{equation}\label{mv}
m_{\nu} =\begin{pmatrix}
y_{1}^2 \Lambda_1+y_{ 2}^2 \Lambda_2+y_{3}^2 \Lambda_3 & a_1 y_{1}^2 \Lambda_1+y_{2}^2 \Lambda_2+a_3 y_{3}^2 \Lambda_3 &a_2 y_{1}^2 \Lambda_1+y_{2}^2 \Lambda_2+a_4 y_{3}^2 \Lambda_3 \\
a_1 y_{1}^2 \Lambda_1+y_{2}^2 \Lambda_2+a_3 y_{3}^2 \Lambda_3 &  a_1^2 y_{1}^2 \Lambda_1+y_{2}^2 \Lambda_2+a_3^2 y_{3}^2 \Lambda_3 & a_1 a_2 y_{1}^2 \Lambda_1+y_{2}^2 \Lambda_2+a_3 a_4 y_{3}^2 \Lambda_3 \\
a_2 y_{1}^2 \Lambda_1+y_{2}^2 \Lambda_2+a_4 y_{3}^2 \Lambda_3 &  a_1 a_2 y_{1}^2 \Lambda_1+y_{2}^2 \Lambda_2+a_3 a_4 y_{3}^2 \Lambda_3 & a_2^2 y_{1}^2 \Lambda_1+y_{2}^2 \Lambda_2+a_4^2 y_{3}^2 \Lambda_3\\
\end{pmatrix}.
\end{equation}
By using the properties  $a_1+a_2=-1$, $a_3+a_4=-1$, $a_1+a_1^2+a_1 a_2=0$, $a_2+a_2^2+a_1 a_2=0$, $a_3+a_3^2+a_3 a_4=0$ and $a_4+a_4^2+a_3 a_4=0$ (see Eqns. (\ref{a1}) to (\ref{a4})), we can clearly see that this matrix is magic \textit{i.e.} the sum of the elements in each row and column is equal to $c_2  y_{ 2}^2 \Lambda_2$ \textit{i.e.}, $m_2$.

\noindent In the upcoming sections (Sections \ref{s5} and \ref{s6}), we shall explore two scenarios. In the first scenario (Section \ref{s5}), we will investigate the implications of trimaximal mixing matrix on both LFV process ($\mu\rightarrow e\gamma$) and the relic density of CDM while simultaneously satisfying neutrino oscillation data, within Scotogenic model of neutrino mass generation. 
In the second scenario (Section \ref{s6}), motivated from our earlier work\cite{Singh:2022nmk}, we shall introduce additional constraint on $m_\nu$,  enabling us to determine independent Yukawa coupling ($y_1$ $(y_3)$ for NH (IH)).


\begin{figure}[t]
	\centering
	\begin{tabular}{cc} 
		\includegraphics[width=0.45\linewidth]{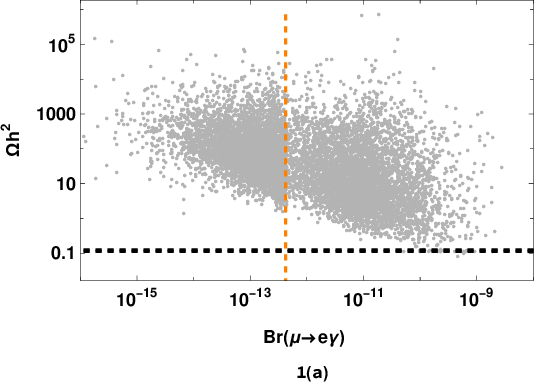}& \includegraphics[width=0.45\linewidth]{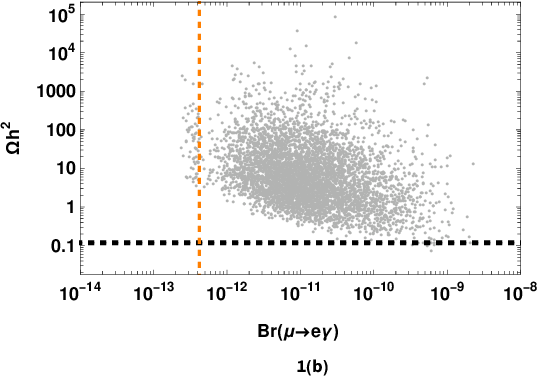}\\
		\includegraphics[width=0.45\linewidth]{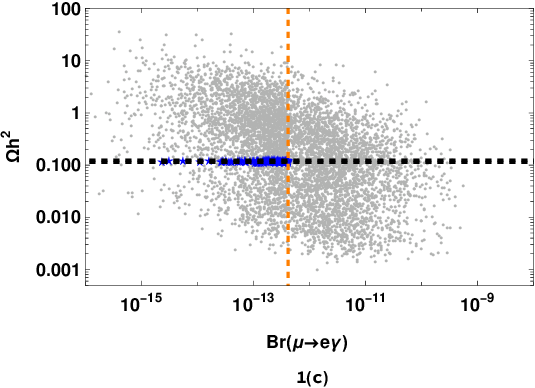}& \includegraphics[width=0.45\linewidth]{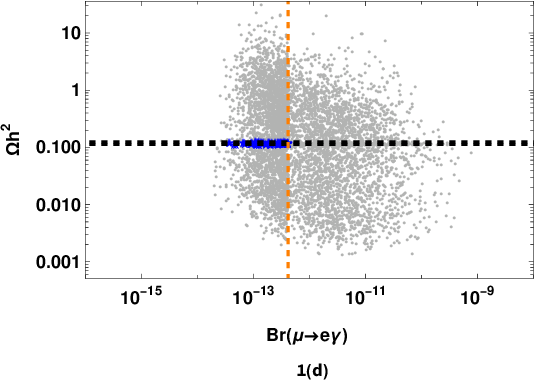}\\
		\includegraphics[width=0.45\linewidth]{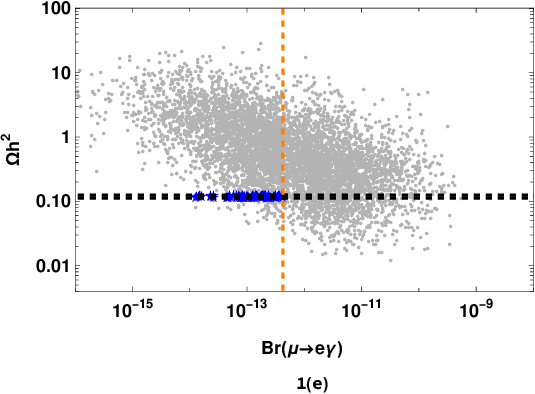}& \includegraphics[width=0.45\linewidth]{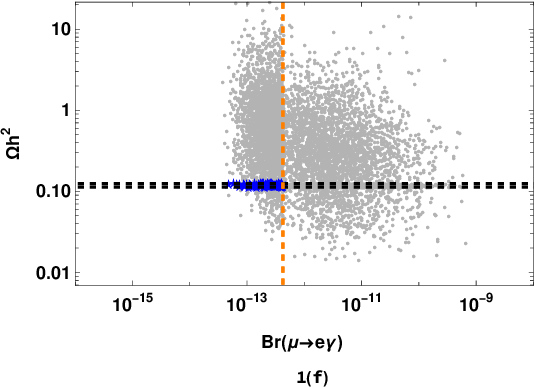}
	\end{tabular}
	\caption{The first, second and third rows represent plots for cases (1), (2) and (3), respectively. First (second) column is for NH (IH) of neutrino masses. The grey points show the parameter space that satisfies neutrino oscillation data at $3\sigma$, while the points shown in blue color  (``\textcolor{blue}{$\filledstar$}") satisfy the \textit{simultaneity condition}. The horizontal  and verticle lines  are experimental range of $\Omega h^2$ and experimental upper bound on Br($\mu\rightarrow e \gamma$) respectively.}
	\label{fig1new}
\end{figure}

 \begin{figure}[t]
	\centering
	\begin{tabular}{cc}
		\includegraphics[width=0.45\linewidth]{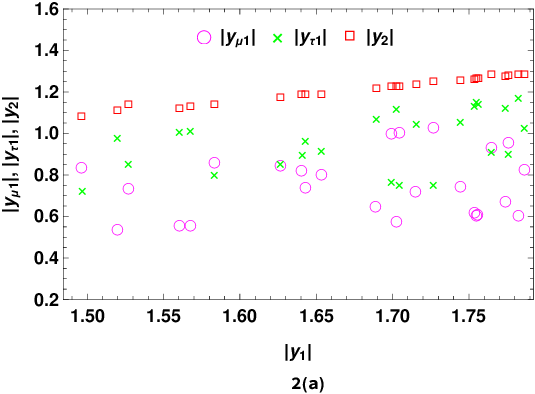}& \includegraphics[width=0.45\linewidth]{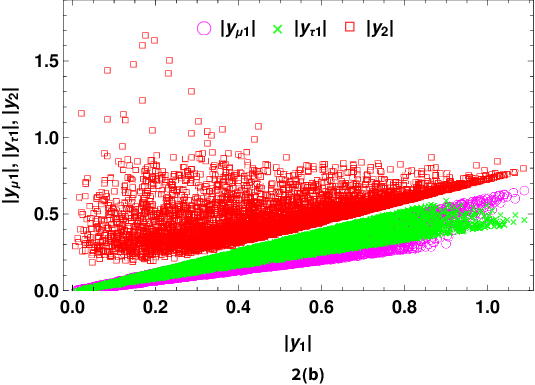}\\
	\includegraphics[width=0.45\linewidth]{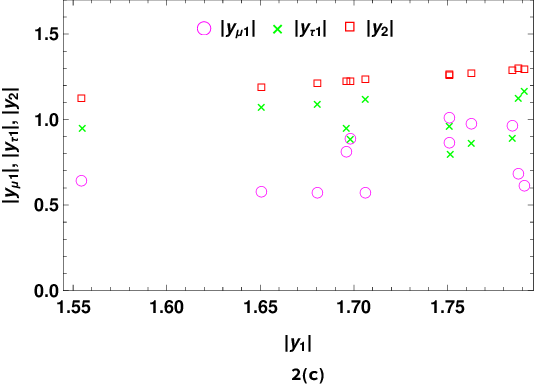}& \includegraphics[width=0.45\linewidth]{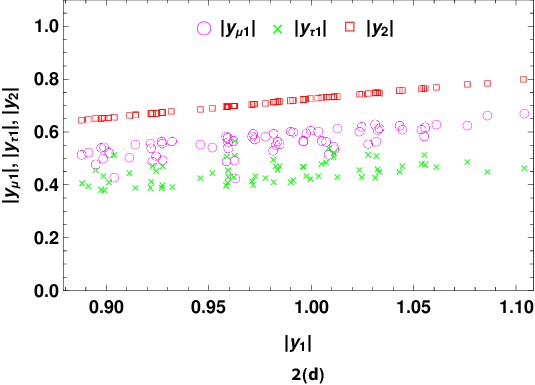}\end{tabular}
	\caption{Case (1): The first (second) column shows Yukawa couplings that satisfy CDM relic density bounds (Yukawa couplings that satisfy the experimental upper bound on Br($\mu \rightarrow e \gamma$) ) for case (1). The first (second) row is for NH (IH) of neutrino masses.}
	\label{fig2new}
\end{figure}

 \begin{figure}[t]
	\centering
	\begin{tabular}{cc}
		\includegraphics[width=0.45\linewidth]{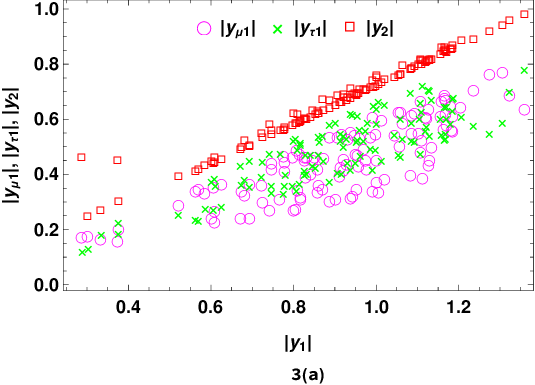}& \includegraphics[width=0.45\linewidth]{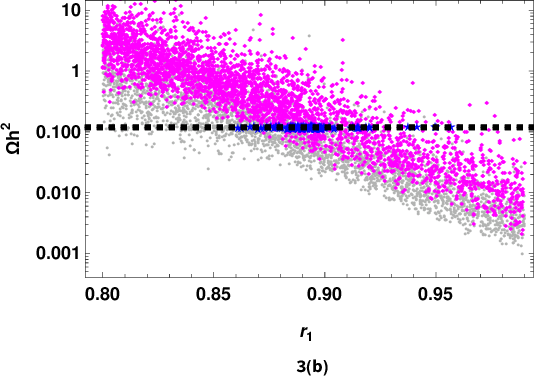}\\
		\includegraphics[width=0.45\linewidth]{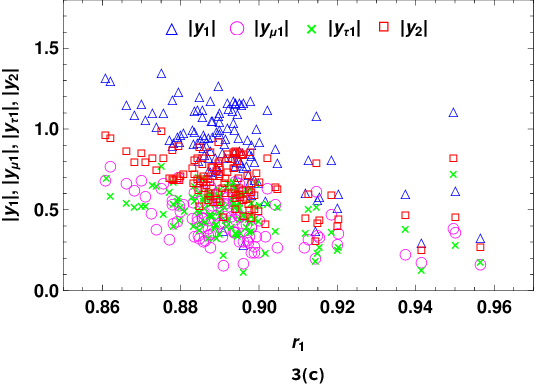}& \includegraphics[width=0.45\linewidth]{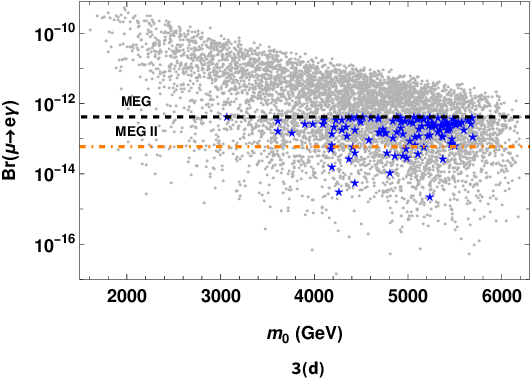}\\
		 \includegraphics[width=0.45\linewidth]{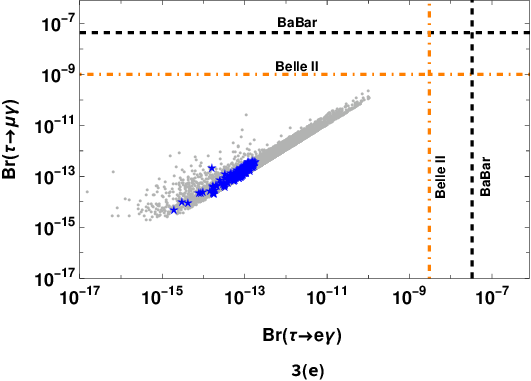}&\includegraphics[width=0.45\linewidth]{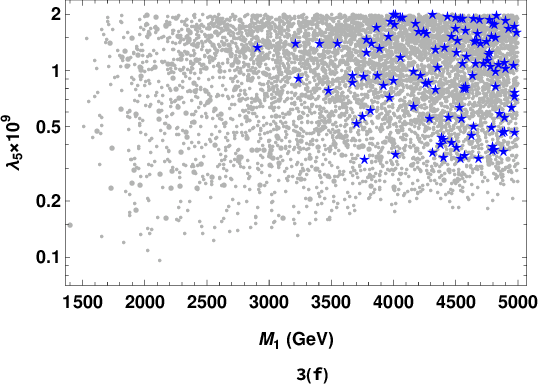}
	\end{tabular}
	\caption{Case (2) for NH: Figure \ref{fig3new}(a) and \ref{fig3new}(c) show the Yukawa coupling strength for points satisfying the \textit{simultaneity condition}. In Fig. \ref{fig3new}(b), diamond-shaped magenta points indicate points satisfying the experimental upper bound on Br($\mu \rightarrow e \gamma$), and black dashed horizontal lines represent the experimentally allowed 3$\sigma$ region for $\Omega h^2$. In Fig. \ref{fig3new}(d) and \ref{fig3new}(e), the black dashed and orange dotted-dashed lines show the sensitivities of the current and future  experiments, respectively. }
	\label{fig3new}
\end{figure}

 \begin{figure}[t]
	\centering
	\begin{tabular}{cc}
		\includegraphics[width=0.45\linewidth]{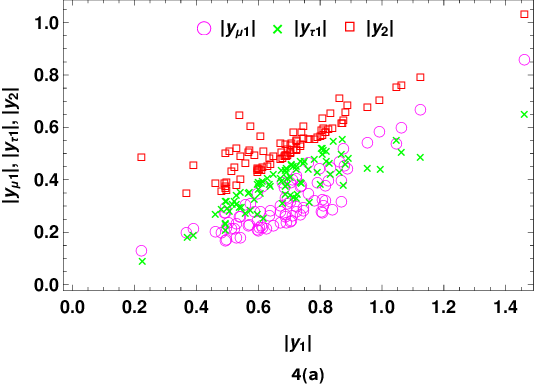}& \includegraphics[width=0.45\linewidth]{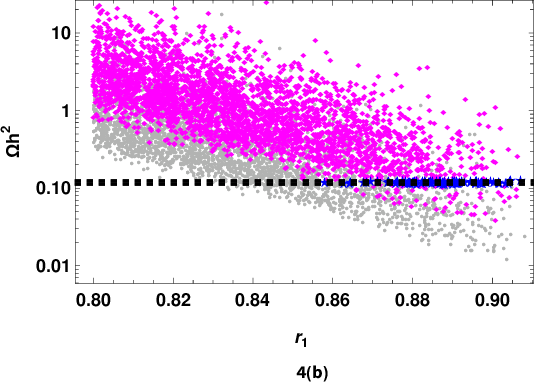}\\
		\includegraphics[width=0.45\linewidth]{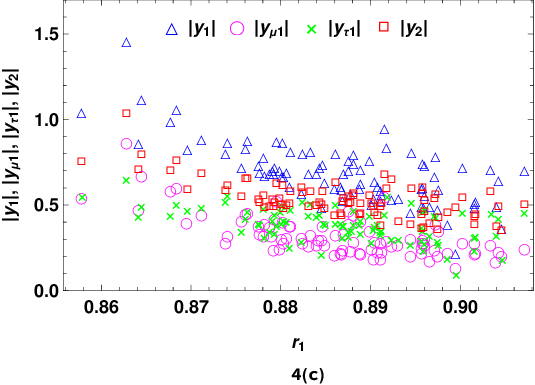}& \includegraphics[width=0.45\linewidth]{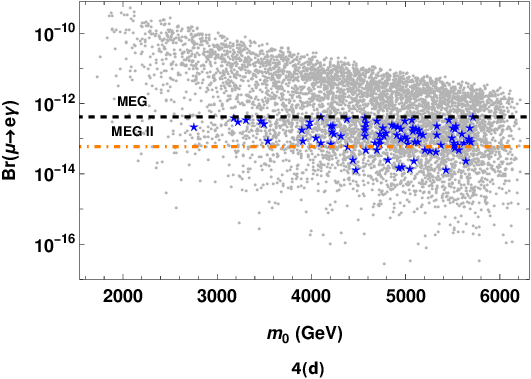}\\
		\includegraphics[width=0.45\linewidth]{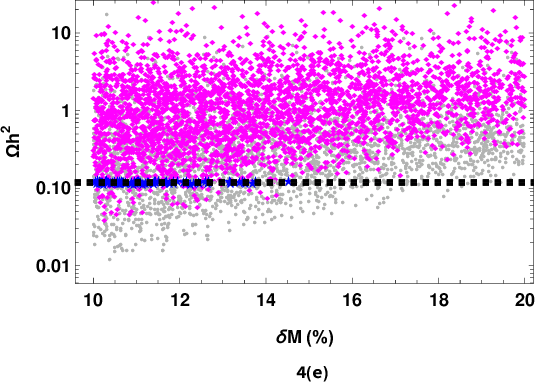}& \includegraphics[width=0.45\linewidth]{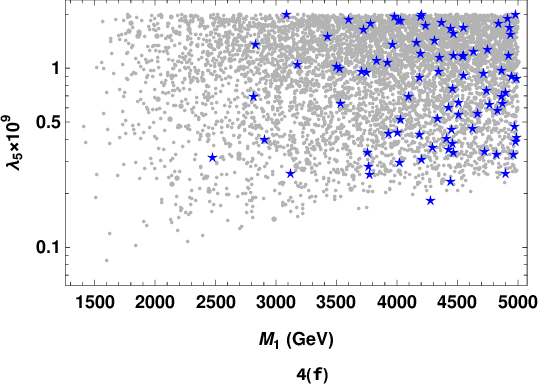}
	\end{tabular}
	\caption{Case (3) for NH: Figure \ref{fig4new}(a) and \ref{fig4new}(c) show the Yukawa coupling strength for points satisfying the \textit{simultaneity condition}. In Fig. \ref{fig4new}(b) and \ref{fig4new}(e), diamond-shaped magenta points indicate points satisfying the experimental upper bound on Br($\mu \rightarrow e \gamma$), and black dashed horizontal lines represent the experimentally allowed 3$\sigma$ region for $\Omega h^2$. In Fig. \ref{fig4new}(d), the black dashed and orange dotted-dashed lines show the sensitivities of the current and future  experiments, respectively. }
	\label{fig4new}
\end{figure}

\section{Numerical Analysis and Discussion}{\label{s5}}

In the previous section, we have determined the structure of the Yukawa coupling matrix $y$ and the neutrino mass matrix $m_{\nu}$ emanating from TM$_2$ mixing paradigm. In this section, we will numerically investigate the neutrino phenomenology arising from this model and its implications for the relic density of cold dark matter ($\Omega h^2$) and possible LFV considering branching ratio of $\mu\rightarrow e \gamma$ process (Br($\mu\rightarrow e \gamma$)). For this study, we vary the neutrino mass-squared differences within their $3\sigma$ range \cite{deSalas:2020pgw}, \textit{viz.} $\Delta m_{21}^2 = (6.94\,\,-\,\,8.14)\times 10^{-5} \text{eV}^2$ and $\Delta m_{31}^2 = (2.47\,\,-\,\,2.63)\times 10^{-3} \text{eV}^2$ for normal hierarchy (NH), and $\Delta m_{31}^2 = (2.37\,\,-\,\,2.53)\times 10^{-3} \text{eV}^2$ for inverted hierarchy (IH) (the negative sign is already included in our analysis, so $\Delta m_{31}^2$ for IH is taken as positive). 

For the parameters originating from the Scotogenic model, we must ensure that $|\lambda_5|<<1$ for small neutrino masses, and for our choice of fermionic dark matter, $N_1$ will be the lightest. As, discussed in Section \ref{s3}, we have considered three cases to study coannihilation dynamics in the model.  The assumptions in each case are:  In case $(1)$ $N_1$ and $N_2$ are nearly degenerate with splitting  $\delta M=(M_2-M_1)/M_1\simeq 0$,  $r_1\leq0.42$ and $r_3>1.3$. In case $(2)$ in addition to $\delta M\simeq 0$ we have $N_1$ and $\zeta$ near in mass so, as a representative manifestation, we set $0.80\leq r_1<0.99$ while $r_3>1.3$. In case $(3)$ we consider non-zero mass splitting, $\delta M\neq 0$, and  choose $\delta M$ to lie in between $(10\%- 20\%)$ range, while other things are kept same as in case $(2)$. Taking these into account, we randomly vary these parameters with a uniform distribution within the specified ranges, as shown in Table \ref{tab1}. In order to  numerically calculate the LFV observables and the CDM relic density, we have used the SARAH-4.15.2 \cite{Staub:2008uz,Staub:2013tta,Staub:2015kfa,Porod:2014xia} generated modules for SPheno-4.0.5 \cite{Porod:2003um,Porod:2011nf} and micrOMEGAs-5.3.41 \cite{Belanger:2001fz,Belanger:2020gnr,Belanger:2021smw,Alguero:2022inz,Belanger:2014vza}, respectively. Additionally, the free parameters $\theta$ and $\phi$, parametrizing TM$_2$ mixing matrix, are randomly varied in the range $0$ to $2\pi$ with uniform distribution.

Before proceeding further, let us write down some important relations used to obtain the physical observable parameters in the neutrino sector. In term of the elements of TM$_2$ mixing matrix (Eqn. (\ref{tm21})), neutrino mixing angles are obtained as 
\begin{equation}
\begin{rcases}
\begin{aligned}
    &\sin^2\theta_{12}=\frac{|(U_{\text{TM}_2})_{12}|^2}{1-|(U_{\text{TM}_2})_{13}|^2}=\frac{1}{\cos 2 \theta+2},\\
    &\sin^2\theta_{13}=|(U_{\text{TM}_2})_{13}|^2=
            \frac{2 \sin ^2\theta}{3},\\ 
    &\sin^2\theta_{23}=\frac{|(U_{\text{TM}_2})_{23}|^2}{1-|(U_{\text{TM}_2})_{13}|^2}=
            \frac{1}{2} \left(\frac{\sqrt{3} \sin 2 \theta  \cos \phi }{\cos 2 \theta +2}+1\right).   
\end{aligned}
\label{mix}
\end{rcases}
\end{equation}
In our analysis we have considered Yukawa coupling matrix $y$ to be complex. Hence using  Eqn. (\ref{m1m2m3_2}), three Majorana phases can be obtained as
\begin{equation}\label{}
   \alpha_1 = \frac{1}{2}\text{arg}[m_1]  ,\,\,\  \alpha_2 = \frac{1}{2}\text{arg}[m_2], \,\,\ \alpha_3 = \frac{1}{2}\text{arg}[m_3] .
\end{equation}
Using these relations, we obtain two physical Majorana phases as 
\begin{equation}
   \alpha_{21}  =\alpha_2-\alpha_1,\,\,\ \alpha_{31}  =\alpha_3-\alpha_1. 
\end{equation}

The Jarlskog $CP$ invariant\cite{Jarlskog:1985ht,Bilenky:1987ty, Krastev:1988yu}, Dirac phase $\delta$, and other two invariants corresponding to Majorana phases $\alpha_{21}$ and $\alpha_{31}$ \cite{Nieves:1987pp,Aguilar-Saavedra:2000jom,Nieves:2001fc,Bilenky:2001rz} are given by 
\begin{eqnarray}
\begin{rcases}
&J_{CP}= \text{Im}[(U_{\text{TM}_2})_{11}(U_{\text{TM}_2})_{12}^*(U_{\text{TM}_2})_{21}^*(U_{\text{TM}_2})_{22}]= \frac{1}{6\sqrt{3}}\sin 2\theta \sin\phi, \\
&\delta= \text{arc}\sin\left[\frac{J_{CP}}{\sin\theta_{12}\cos\theta_{12} \sin\theta_{23} \cos\theta_{23}\sin\theta_{13}\cos^2\theta_{13}}\right],\\
&I_{1}=\text{Im}[(U_{\text{TM}_2})_{11}^*(U_{\text{TM}_2})_{12}e^{-i\alpha_{21}}]=-\frac{\sqrt{2}}{3}\cos\theta\sin \alpha_{21}, \\ 
&I_{2}=\text{Im}[(U_{\text{TM}_2})_{11}^*(U_{\text{TM}_2})_{13}e^{-i\alpha_{31}}]=-\frac{1}{3}\sin 2\theta \sin \alpha_{31}. 
\end{rcases}
\label{inv}
\end{eqnarray}

The effective Majorana mass, denoted as $|m_{ee}|=|(m_v)_{11}|=\left|\sum_{i=1}^{3}(U_{\text{TM}_2})^2_{1i} m_i\right|$, is a crucial observable in  neutrinoless double beta decay ($0 \nu \beta \beta$) experiments, which aim to establish the Majorana nature of neutrinos.

\begin{figure}[t]
  \centering
\begin{tabular}{cc} 
	\includegraphics[width=0.45\linewidth]{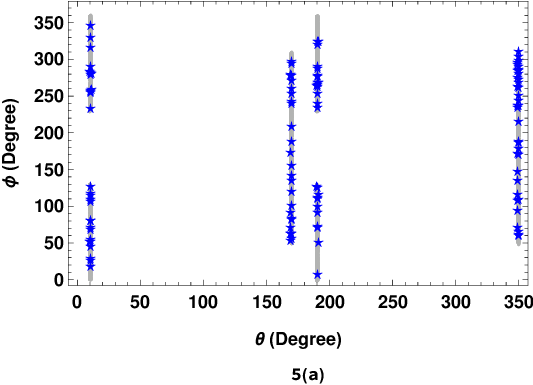}& \includegraphics[width=0.45\linewidth]{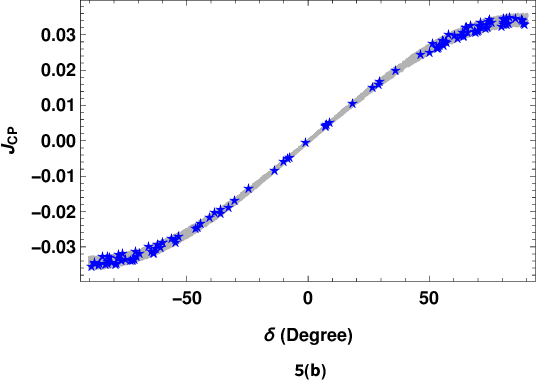}\\ 
	\includegraphics[width=0.45\linewidth]{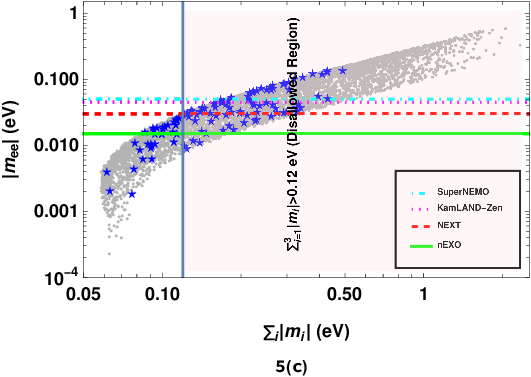}& \includegraphics[width=0.45\linewidth]{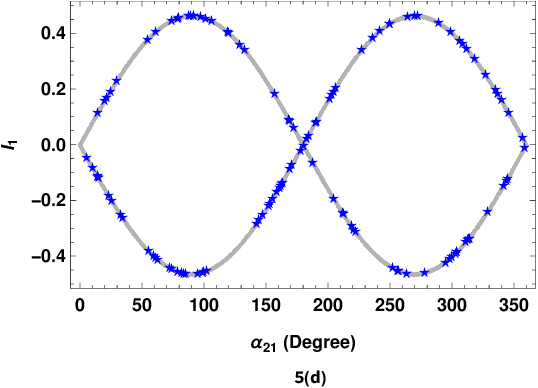}\\ 
	\includegraphics[width=0.45\linewidth]{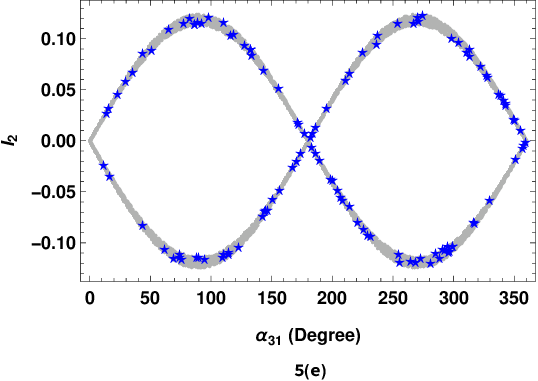}& \includegraphics[width=0.45\linewidth]{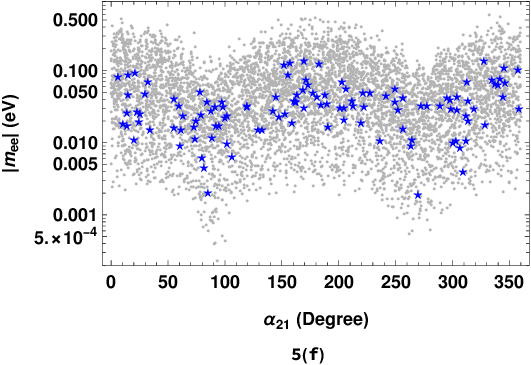}\\ 
	\includegraphics[width=0.45\linewidth]{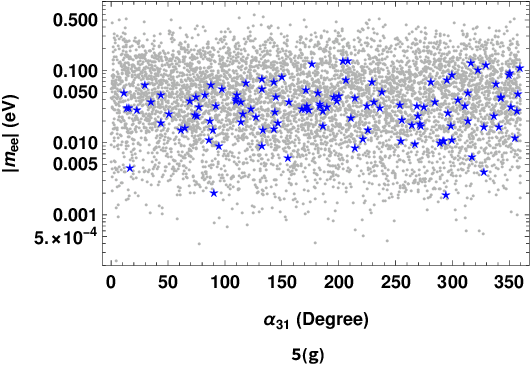}& \includegraphics[width=0.45\linewidth]{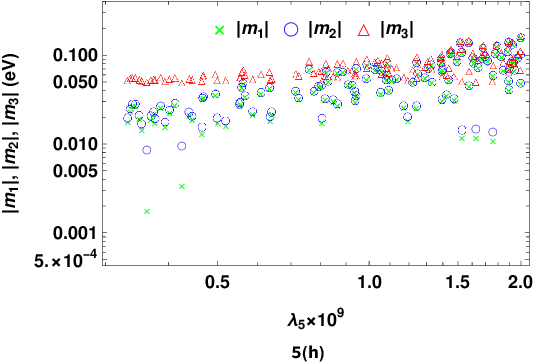}\\ \end{tabular}
   \caption{Case (2) for NH:  The grey points show the parameter space that satisfies neutrino oscillation data at $3\sigma$,  while the points shown in blue color  (``\textcolor{blue}{$\filledstar$}") satisfy the \textit{simultaneity condition}. In Fig. \ref{fig1}(h) all points satisfy \textit{simultaneity condition}.}
  \label{fig1}
\end{figure}

 \begin{figure}[t]
	\centering
	\begin{tabular}{cc}
		\includegraphics[width=0.45\linewidth]{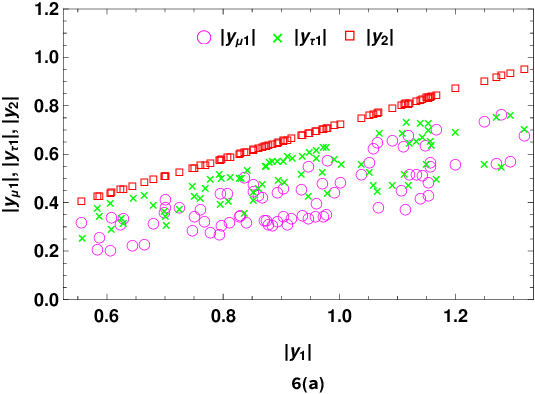}& \includegraphics[width=0.45\linewidth]{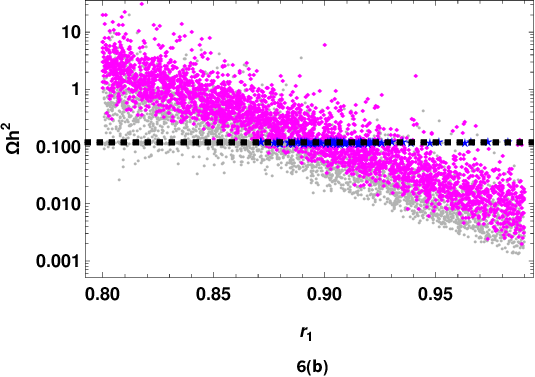}\\
		\includegraphics[width=0.45\linewidth]{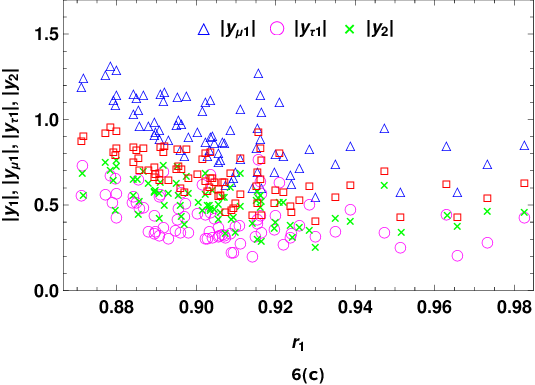}& \includegraphics[width=0.45\linewidth]{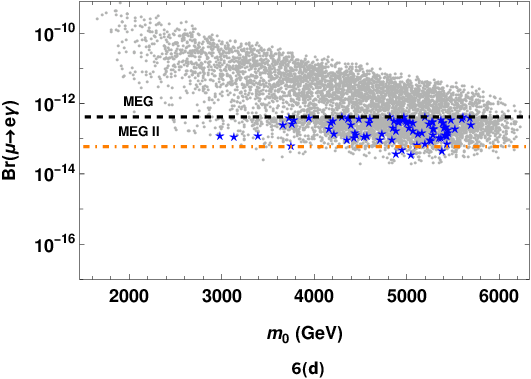}\\
		\includegraphics[width=0.45\linewidth]{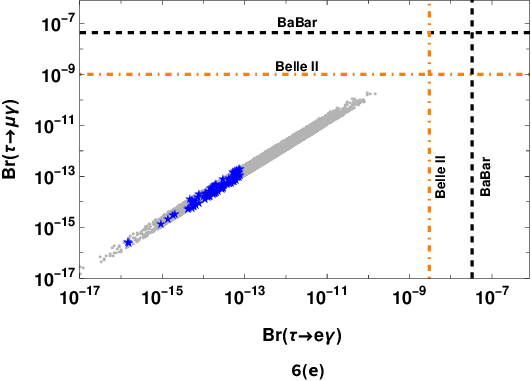}&
		\includegraphics[width=0.45\linewidth]{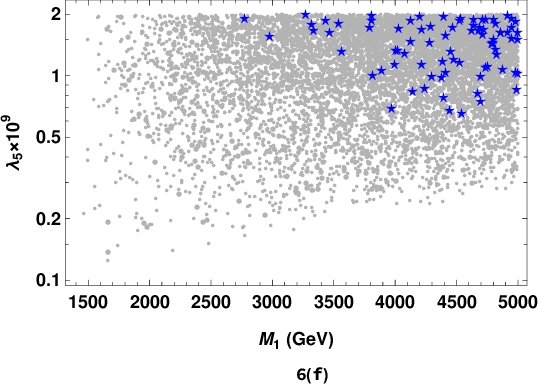}
	\end{tabular}
	\caption{Case (2) for IH: Figure \ref{fig6new}(a) and \ref{fig6new}(c) show the Yukawa coupling strength for points satisfying the \textit{simultaneity condition}. In Fig. \ref{fig6new}(b), diamond-shaped magenta points indicate points satisfying the experimental upper bound on Br($\mu \rightarrow e \gamma$), and black dashed horizontal lines represent the experimentally allowed 3$\sigma$ region for $\Omega h^2$. In Fig. \ref{fig6new}(d) and \ref{fig6new}(e), the black dashed and orange dotted-dashed lines show the sensitivities of the current and future  experiments, respectively.}
	\label{fig6new}
\end{figure}

 \begin{figure}[t]
	\centering
	\begin{tabular}{cc}
		\includegraphics[width=0.45\linewidth]{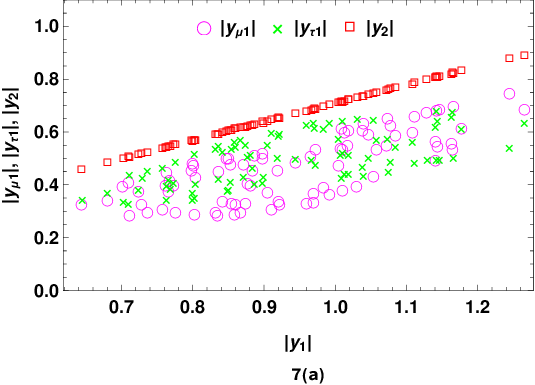}& \includegraphics[width=0.45\linewidth]{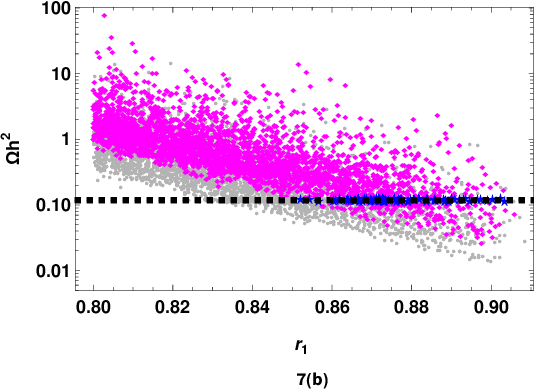}\\
		\includegraphics[width=0.45\linewidth]{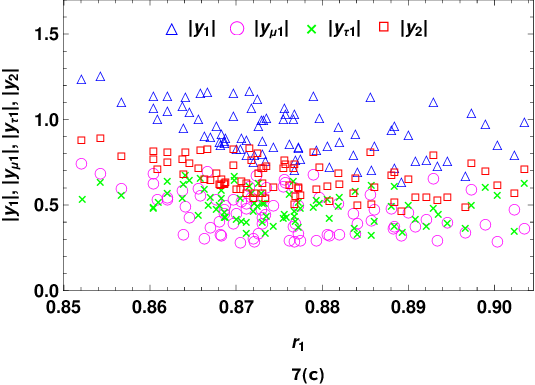}& \includegraphics[width=0.45\linewidth]{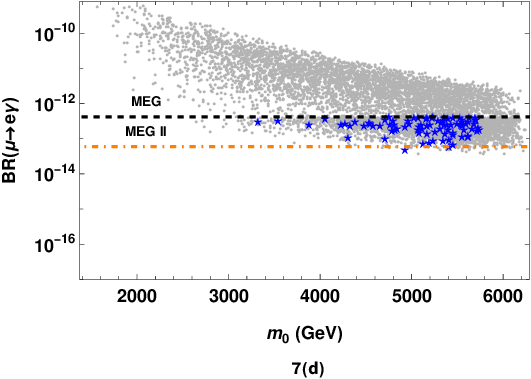}\\
		\includegraphics[width=0.45\linewidth]{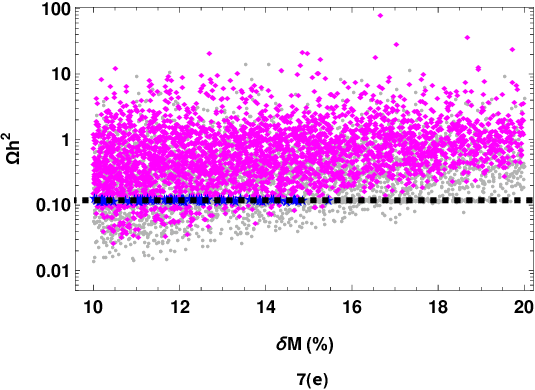}&
		\includegraphics[width=0.45\linewidth]{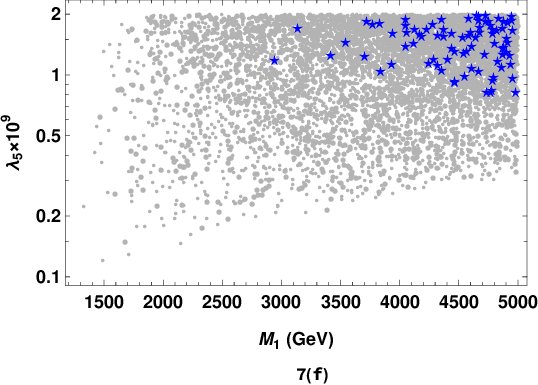}
	\end{tabular}
	\caption{Case (3) for IH: Figure \ref{fig7new}(a) and \ref{fig7new}(c) show the Yukawa coupling strength for points satisfying the \textit{simultaneity condition}. In Fig. \ref{fig7new}(b) and \ref{fig7new}(e), diamond-shaped magenta points indicate points satisfying the experimental upper bound on Br($\mu \rightarrow e \gamma$), and black dashed horizontal lines represent the experimentally allowed 3$\sigma$ region for $\Omega h^2$. In Fig. \ref{fig7new}(d), the black dashed and orange dotted-dashed lines show the sensitivities of the current and future  experiments, respectively.}
	\label{fig7new}
\end{figure}


\begin{figure}[t]
  \centering
\begin{tabular}{cc} 
\includegraphics[width=0.45\linewidth]{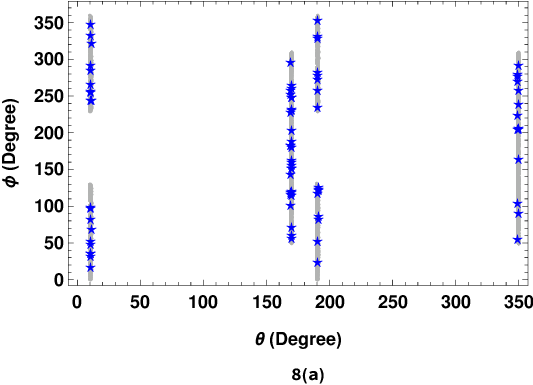}& \includegraphics[width=0.45\linewidth]{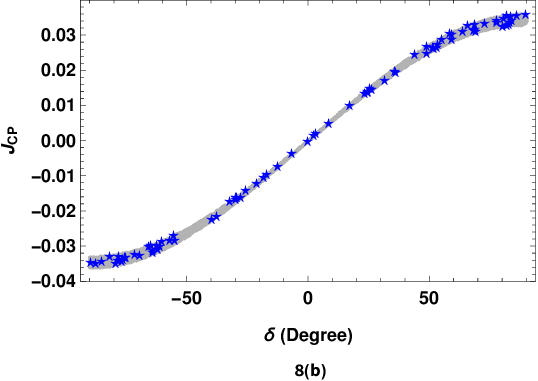}\\ 
 \includegraphics[width=0.45\linewidth]{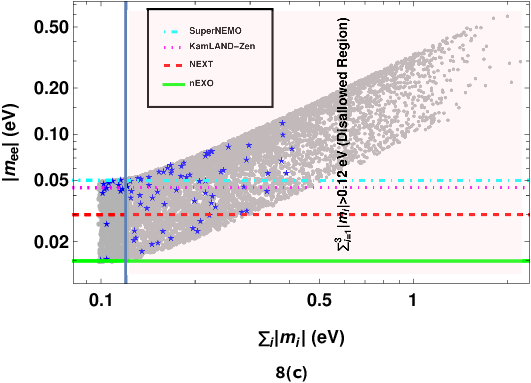}& \includegraphics[width=0.45\linewidth]{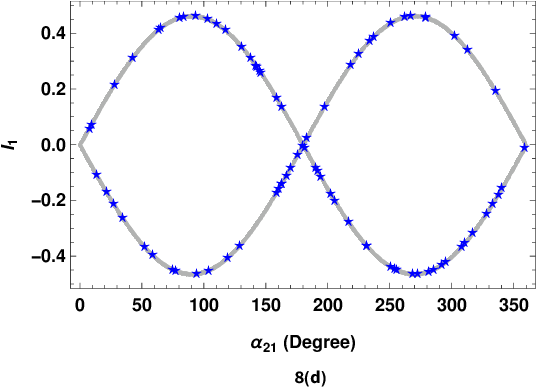}\\ 
 \includegraphics[width=0.45\linewidth]{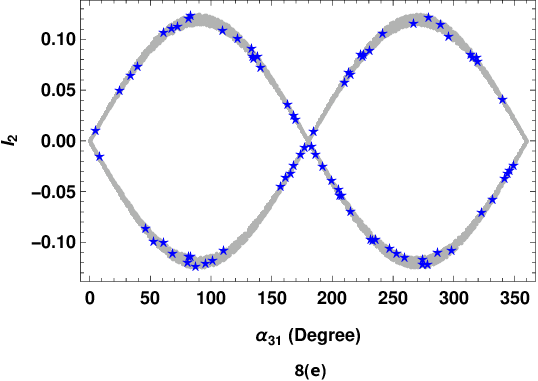}& \includegraphics[width=0.45\linewidth]{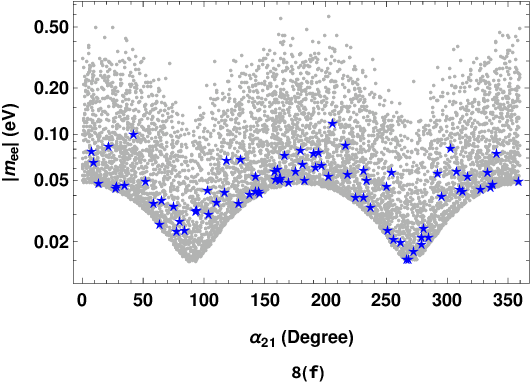}\\ 
  \includegraphics[width=0.45\linewidth]{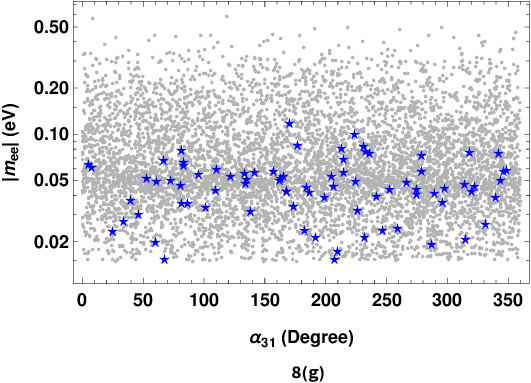}& \includegraphics[width=0.45\linewidth]{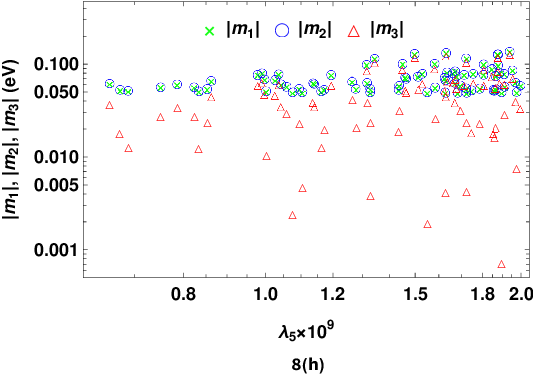}\\ \end{tabular}
 \caption{Case (2) for IH:  The grey points show the parameter space that satisfies neutrino oscillation data at $3\sigma$,  while the points shown in blue color  (``\textcolor{blue}{$\filledstar$}") satisfy the \textit{simultaneity condition}. In Fig. \ref{fig3}(h) all points satisfy \textit{simultaneity condition}.}
  \label{fig3}
\end{figure}

\begin{table}[t]
	\begin{center}
		\begin{tabular}{|l|l|}
			\hline
			$\lambda_2$&  $[ 10^{-5},\;1]$\\
			\hline
			$\lambda_3$& $[ 10^{-2},\;1]$\\ 
			\hline
			$\lambda_4$& $[ 10^{-8},\;10^{-7}]$\\
			\hline
			$\lambda_5$& $[10^{-11},\;2\times 10^{-9}]$\\
			\hline
			$m_{\zeta}$ (GeV)& $[10^3,\;6.5\times 10^3]$\\
			\hline
			$M_1$ (GeV)& $[1,\;5000]$\\
			\hline
			$M_2$ (GeV)& $[1,\;5000]$\\
			\hline
		    $M_3$ (GeV)& $[1,\;10000]$\\
			\hline\end{tabular}
	\end{center}
	\caption{\label{tab1}Ranges for parameters used in the numerical analysis.}
\end{table}

\subsection{Normal Hierarchy (NH)}{\label{s5.1}}

For normal hierarchy, we only have one free complex Yukawa coupling, $y_1$, which is varied in the range $0\leq|y_1|\leq 1.8$. The Majorana phases $\alpha_2$ and $\alpha_3$ are freely varied using uniform random numbers in the range $0$ to $2 \pi$. We need to satisfy the neutrino oscillation data \cite{deSalas:2020pgw} and the experimental constraints on the CDM relic density given by $0.1126\leq\Omega h^2\leq0.1246$ \cite{Planck:2018vyg} in their 3$\sigma$ range, as well as the upper bound on the LFV branching ratio, Br($\mu\rightarrow e \gamma$)$\leq 4.2 \times 10^{-13}$ \cite{MEG:2016leq}. We refer to this condition of simultaneous satisfaction as the ``\textit{simultaneity condition}". In  this paper, we present correlation plots mainly with two types of points: the grey points represent parameter space that satisfy the neutrino oscillation data at $3\sigma$, while points denoted by ``\textcolor{blue}{$\filledstar$}", which are of primary interest, satisfy the \textit{simultaneity condition}. In Fig. \ref{fig1new} (first column) we have shown the correlation plot between CDM relic density ($\Omega h^2 $) and LFV branching ratio Br($\mu\rightarrow e \gamma$) for all three cases. \\ 	
  In case (1), where $N_1$ and $N_2$ coannihilations are considered, we find that, \textit{simultaneity condition} cannot be satisfied. The experimental upper bound on branching ratio Br$(\mu \rightarrow e \gamma)$ is satisfied for the values where CDM relic density is well above  experimentally observed abundance (see Fig. \ref{fig1new} (a)). Now, from Eqn. (\ref{yu2}), consider the Yukawa couplings $|y_{e 1}|=|y_1|$, $|y_{\mu 1}|=|a_1 y_1|$, $|y_{\tau 1}|=|a_2 y_1|$, and $|y_{e 2}|=|y_{\mu 2}|=|y_{\tau 2}|=|y_2|$ responsible for the $N_1$ and $N_2$ coannihilation processes. In Fig. \ref{fig2new}, these Yukawa couplings are shown as a function of  coupling $|y_1|$  for cases when CDM relic density constraint is satisfied (Fig. \ref{fig2new}(a)) and when LFV branching ratio constraint for $\mu \rightarrow e \gamma $ is satisfied (Fig. \ref{fig2new}(b)). It is evident from these plots that CDM relic density is satisfied for higher values of Yukawa couplings with $ |y_1|\gtrsim 1.49$ whereas Br($\mu \rightarrow e \gamma$) requires lower values of these couplings with $|y_1|\lesssim 1.1$. As there is no  common parameter space available, so \textit{simultaneity condition} cannot be satisfied in this case. In Fig. \ref{fig2new}(a) there is a particular hierarchy among these Yukawa couplings  i.e., $|y_{\mu 1}|,|y_{\tau 1}|< |y_2|< |y_1|$. However, hierarchy between $|y_{\mu 1}|$ and $|y_{\tau 1}|$ is not fixed and can change over for  different values of $|y_1|$. In this parameter space,  $|y_1|= |y_{e1}|$ is largest so we will observe annihilation processes where $N_1$ annihilates mainly to $e^-e^+$ whereas $N_2$ can annihilate to any of charged lepton anti-lepton pairs because of flavour structure (Eqn. \ref{yu2}) its coupling with all charged leptons has same strength. The coannihilation processes between $N_1$ and $N_2$ gives  charged lepton and anti-lepton pairs in final state.
 \\
In case (2), coannihilation among $N_1$, $N_2$ and $\zeta$ is considered which can satisfy the \textit{simultaneity condition}  as evident from Fig. \ref{fig1new}(c). The strength of the Yukawa couplings corresponding to $N_1$ and $N_2$ is shown in Fig. \ref{fig3new}(a) for points which satisfy the \textit{simultaneity condition}. Here the  Yukawa coupling $|y_1|$ is varying in the range $0.3\lesssim |y_1|\lesssim 1.35 $. There exist hierarchy among Yukawa couplings i.e., $|y_{\mu 1}|, \;|y_{\tau 1}|<|y_2|<|y_1|$, for value of $|y_1|> 0.5$ however hierarchy among $|y_{\mu 1}|$ and $|y_{\tau 1}|$ is not  same as it changes with value of $|y_1|$. For values of $|y_1|$ lower than $0.5$, it is clear that $|y_{\mu 1}|,|y_{\mu 1}|<|y_{1}|,|y_{2}|$ but now hierarchy among $|y_1|$ and $|y_2|$ is not fixed.
 Now the Yukawa couplings of $N_1$ and $N_2$ can take smaller values because coannihilation processes has enhanced the effective cross-section. In Fig. \ref{fig3new}(b)  we have shown the CDM relic density ($\Omega h^2 $) and $r_1$ correlation plot where diamond shaped magenta points are showing the region  where  Br($\mu \rightarrow e \gamma$)$\leq4.2\times 10^{-13}$ constraint is satisfied. The black dotted horizontal lines are the experimentally allowed 3$\sigma$ region, $0.1126\leq\Omega h^2\leq0.1246$.  It is  evident from the figure that for  $r_1<0.86$  the branching ratio constraint is satisfied but CDM relic density is over abundant because in that region the Yukawa couplings can take, relatively small values. In order to satisfy the \textit{simultaneity condition} we need this band of points to attain successful dark matter relic density which happens only after  right amount  of coannihilation starts for mass ratio $r_1\geq 0.86$. However, over coannihilations results in  under abundant relic density for which \textit{simultaneity condition} cannot be satisfied (see region around $r_1$ near $0.99$). If there were not enough coannihilations then this band of points will never reach the CDM relic density, satisfying region (see region around $r_1$ near $0.80$). In  Fig. \ref{fig3new}(c), we have depicted Yukawa couplings corresponding to $N_1$ and $N_2$ as a function of $r_1$. It is evident from the figure that  as $r_1$ takes on higher values, coannihilation increases with $\zeta$ and Yukawa couplings can now take lower values. In Fig. \ref{fig3new}(d) correlation plot between Br($\mu \rightarrow e \gamma$) and inert scalar mass $m_0$ is shown. The horizontal black dashed  and orange dotted-dashed lines represent the experimental sensitivities of MEG experiment \cite{MEG:2016leq} and future sensitivity of MEG II experiment \cite{MEGII:2018kmf}, respectively. The viable  values of  Br($\mu \rightarrow e \gamma$) can be as small as $\mathcal{O}(10^{-15})$ from which it is evident that MEG II cannot fully probe the viable parameter space for Br($\mu \rightarrow e \gamma$). The \textit{simultaneity condition} keeps value of $m_0$ in the range  $\approx (3000-5700)$ GeV (see Fig. \ref{fig3new}(d)). In Fig. \ref{fig3new}(e) we have shown Br($\tau \rightarrow \mu \gamma$) and Br($\tau \rightarrow e \gamma$)  where horizontal and vertical black dashed lines are showing experimental upper bound from BaBar \cite{BaBar:2009hkt} and orange dotted-dashed lines showing future sensitivities of Belle II \cite{Belle-II:2018jsg} experiment. This figure demonstrate the effectiveness of our approach, which focuses solely on $\mu \rightarrow e \gamma$, in constraining our parameter space alongside the CDM relic density. It is evident that by considering only less stringent constraints, we would fail to tightly constrain the parameter space. Therefore, even if we incorporate these less constraining bounds into our analysis, they would not  affect our parameter space that we have obtained. The region which satisfy the \textit{simultaneity condition} lies well beyond the future experimental sensitives. The parameter $\lambda_5$ which is required to be small in this model, to have small neutrino masses without requiring small Yukawa couplings, takes  values in the restricted region with minimum value around  $3\times 10^{-10}$. For  lower values of $\lambda_5$ the strength of Yukawa couplings can go high, however, since we have restricted our Yukawa couplings in the perturbative limit $|y_{\alpha i}|\leq 1.8$ value of $\lambda_5$ cannot be too small. So, in Fig. \ref{fig3new}(f), emergence of constrained region for $\lambda_5$ is consequence of the interplay between neutrino mass, perturbativity limit and \textit{simultaneity condition}. The dark matter mass lies in the  range $\approx (2900-5000)$ GeV (see Fig. \ref{fig3new}(f)).\\
In case (3), $N_1$, $N_2$ and $\zeta$ coannihilation is studied with non-zero splitting between $N_1$ and $N_2$, $\delta M$ lying in the $(10\% - 20 \%)$ range. In this case, also, \textit{simultaneity condition} can be satisfied but this time the parameter space is more restricted than the without splitting case Fig. \ref{fig1new}(e). In Fig. \ref{fig4new}(a) strength of Yukawa couplings are shown for which the \textit{simultaneity condition} is satisfied. In this case  Yukawa couplings strength of $|y_1|$ varies from around $0.2$ to $1.46$. In this case for $|y_1|> 0.6$ shows same hierarchy as for the without splitting case i.e., $|y_{\mu 1}|, \;|y_{\tau 1}|<|y_2|<|y_1|$ and for $|y_1|\leq 0.6$ no definite hierarchy among $|y_1|$ and $|y_2|$ exist. In Fig.  \ref{fig4new}(b), as an effect of splitting between $N_1$ and $N_2$, in this case CDM relic density experimental constraint is not being satisfied in lower value regions of $r_1$ i.e., around 0.80. As $r_1$ increases and reaches the value $r_1 \gtrsim 0.86$ the band of point satisfying branching ratio constraint for $\mu \rightarrow e \gamma$ process reaches the experimentally allowed region for CDM relic density. This time, $r_1$ is not able to go beyond $0.91$. This is due to the interplay between  hierarchy condition among the coannihilating particles $(M_1< M_2< m_0)$ and the mass splitting condition between $N_1$ and $N_2$. The splitting condition restricts how close $m_0$ can approach $M_1$. The largest value of $r_1$ occurs for smallest value of mass splitting. Due to the minimum  $10\%$  mass splitting constraint,  $r_1$ cannot exceed $0.91$.   Figure \ref{fig4new}(c) shows the variation of Yukawa coupling strength with the mass ratio $r_1$. Closer  masses of $M_1$ and $m_0$, allow the Yukawa couplings to take lower values and hence magenta color band of points can even cross CDM relic density experimental constraint  as shown in  Fig. \ref{fig4new}(b). The mass $m_0$ varies in the range $(2750 -5700)$ GeV (see Fig. \ref{fig4new}(d)).  So, observing from both cases (case (1) and case (2)) we can say that inert scalar mass $m_0$ has upper bound around 5700 GeV. In this case  lowest value for Br($\mu \rightarrow e \gamma$) is $\mathcal{O}(10^{-14})$. Significant part of Br($\mu \rightarrow e \gamma$) parameter space can be probed by future experiment MEG II (see Fig. \ref{fig4new}(d)). The results for the branching ratios of the $\tau \rightarrow e \gamma$ and $\tau \rightarrow \mu \gamma$ processes are not significantly different from the case without splitting, so we have not shown them here. In Fig. \ref{fig4new}(e) we can see that for those points which are satisfying the \textit{simultaneity condition} lying in the range where mass splitting $\delta M$ varies from  $10\% \leq\delta M\leq 14.5\%$. For higher values of splitting,  contribution of coannihilation processes decreases and we have no region where \textit{simultaneity condition} can be satisfied.  The dark matter particle mass can vary from $(2500 - 5000)$ GeV and $\lambda_5$ is restricted in the range $(1.8\times10^{-10}- 2\times10^{-9})$ (see Fig. \ref{fig4new}(f)).

\noindent Now, by considering case (2) we will study neutrino phenomenology.
 We have identified benchmark point corresponding to $y_1=0.245-0.556 i$,  $\alpha_2=236.417^o$ and $\alpha_3=224.963^o$, which,  satisfy the \textit{simultaneity condition}  as shown in third column of Table \ref{tab2}.
The correlation plot between parameters $\theta$ and $\phi$ are shown in Fig. \ref{fig1}(a). It is evident from Fig.  \ref{fig1}(a) that there exist two distinct regions of parameter space under simultaneity constraint \textit{viz.,} (i) for $\theta\backsimeq 10^o \text{ or } 190^o$: $\phi\in (0^o-130^o)\oplus (230^o-360^o)$ (ii) for $\theta\backsimeq 170^o \text{ or } 350^o$: $\phi\in (50^o-310^o)$. The Jarlskog CP invariant varies between the range of $(-0.036$ to $0.036)$, and the Dirac CP-violating phase $\delta$ varies between $(-90^o$ to $90^o)$ approximately. The blue points are  spread over the entire range of these values, as shown in Fig. \ref{fig1}(b).  

The values of the effective Majorana mass $|m_{ee}|$ and the sum of neutrino masses $\sum_{i}|m_i|$ are constrained by the \textit{simultaneity condition}, as depicted in Fig. \ref{fig1}(c).  The values allowed by \textit{simultaneity condition} for $|m_{ee}|$ lie in the range $(0.002$ to $0.15)$ eV, and for $\sum_{i} |m_i|$,  values lie in $(0.05$ to $0.50)$ eV range. The shaded region represents the cosmologically disallowed region from Planck data [TT, TE, EE+lowE+lensing+BAO], which imposes a stringent upper bound of $0.12$ eV at $95\%$ confidence level (CL) \cite{Planck:2018vyg} for the sum of neutrino masses $(\sum_{i=1}^3|m_i|)$. The horizontal lines indicate the experimental sensitivity of current and future $0\nu \beta \beta$ decay experiments: SuperNEMO\cite{Barabash:2011row}, KamLAND-Zen\cite{KamLAND-Zen:2016pfg}, NEXT\cite{NEXT:2009vsd,NEXT:2013wsz}, and nEXO\cite{Licciardi:2017oqg}. Considering the cosmological bound on sum of neutrino masses $\sum_{i} |m_i|$, the \textit{simultaneity condition} predicts $|m_{ee}|\in (0.002-0.03)$ eV range, some region of which can be probed at nEXO $0\nu \beta \beta$ decay experiment. However, to reach a final conclusion, the sensitivity of these $0\nu \beta \beta$ decay experiments needs to be improved by an order of magnitude factor.

In addition to the Jarlskog CP invariant, the Majorana CP invariants $I_1$ and $I_2$ are also computed with Majorana phases $\alpha_{21}$ and $\alpha_{31}$, as shown in Figs. \ref{fig1}(d) and \ref{fig1}(e) respectively. For $I_1$, the allowed range is $(-0.46 \,\,\text{to}\,\, 0.46)$, and for $I_2$, it is $(-0.12 \,\,\text{to}\,\, 0.12)$. Both $\alpha_{21}$ and $\alpha_{31}$ vary in the entire range from $0^o$ to $360^o$.

In Figs. \ref{fig1}(f) and \ref{fig1}(g), we depict the correlation plot of Majorana phases $\alpha_{21}$ and $\alpha_{31}$ with $|m_{ee}|$, respectively. It is evident from Fig. \ref{fig1}(f) that $|m_{ee}|$ attains maximum value when the Majorana CP phase $\alpha_{21}$ is near $0^o,\,180^o\, \text{or} \,360^o$, and minimum values around $90^o \,\text{or}\, \,270^o$. However, there is no sharp correlation visible with $\alpha_{31}$. 

In Fig. \ref{fig1}(h), we present the plot of the Higgs quartic coupling $\lambda_5$ and three neutrino masses, where each point, shown in different colors and shapes, satisfies the \textit{simultaneity condition}. The smallest value of the lightest neutrino mass $|m_1|$ is $0.002$ eV which corresponds to the  value of $\lambda_5\,\, (\approx 3.6 \times 10^{-10}$). It can be observed that for smaller values of $\lambda_5$, the neutrino masses can also be small. As $\lambda_5$ increases, neutrino masses, also, increase and tend towards a degenerate region.

\subsection{Inverted Hierarchy (IH)}

For inverted hierarchy, the free complex Yukawa coupling is, $y_3$, which is varied in the range $0\leq|y_3|\leq 1.8$. The Majorana phases $\alpha_1$ and $\alpha_2$ are randomly varied using uniform distribution in the range $0$ to $2 \pi$. In Fig. \ref{fig1new} (second column) we have shown the correlation plot between CDM relic density and LFV branching ratio Br($\mu\rightarrow e \gamma$) for all three  coannihilation cases. As we have seen for NH, it is clear that in case (1) the \textit{simultaneity condition} cannot be satisfied (see Fig. \ref{fig1new} (b)). The reason for this is similar to as discussed in NH case which can be understood by looking at Yukawa coupling strengths of $N_1$ ($|y_{\alpha 1} |$) and $N_2$ ($|y_{\alpha 2}|$) as shown in Fig. \ref{fig2new}(c) for points satisfying CDM relic density experimental constraint and Fig. \ref{fig2new}(d) for points satisfying the experimental upper bound for Br$(\mu\rightarrow e \gamma)$. The comparison between these two shows us that Yukawa couplings strength for both is very different for CDM relic density $|y_1| \gtrsim 1.55 $ and for Br($\mu \rightarrow e \gamma$) we have $|y_1|\lesssim 1.1$. As there is no common parameter space in these Yukawa couplings hence \textit{simultaneity condition} cannot be satisfied. Annihilation and coannihilation processes will remain the same as discussed in NH case because here also exist the same hierarchy  among Yukawa couplings.\\
	 In case (2) \textit{simultaneity condition} can be satisfied as evident from Fig. \ref{fig1new}(d). The corresponding strength of Yukawa couplings which satisfy the  \textit{simultaneity condition} is shown in Fig. \ref{fig6new}(a). In whole allowed parameter space there exist hierarchy among Yukawa couplings i.e., $|y_{\mu 1}|, \;|y_{\tau 1}|<|y_2|<|y_1|$, however hierarchy among $|y_{\mu 1}|$ and $|y_{\tau 1}|$ is not same and changes with value of $|y_1|$. We can  infer from the plot that for majority of points, $|y_{\mu 1}|<|y_{\tau 1}|$. It is evident from, CDM relic density and $r_1$ correlation plot,  shown in Fig. \ref{fig6new}(b), that the \textit{simultaneity condition} is satisfied for $r_1\gtrsim 0.87$. Figure  \ref{fig6new}(c) shows strength of Yukawa couplings as a function of $r_1$, more higher the $r_1$ more smaller values the Yukawa couplings can take. As shown in Fig. \ref{fig6new}(d) the mass of inert scalar is constrained by \textit{simultaneity condition} in the range $(3000 - 5700)$ GeV. The MEG II experiment is capable of testing most of part of Br($\mu \rightarrow e \gamma$) parameter space as lowest value is $\mathcal{O}(10^{-14})$. Again it is evident from Fig. \ref{fig6new}(e) that the constraint for Br($\tau \rightarrow \mu \gamma$) and Br($\tau \rightarrow e \gamma$) are not enough to constrain the parameter space and even future experiment cannot test predictions for parameter space obtained from this model. The dark matter mass varies from $(2750 - 5000)$ GeV and Higgs quartic coupling  $\lambda_5$  varies from $(6.5\times 10^{-10}-2\times10^{-9})$ (see Fig. \ref{fig6new}(f)).
	  \\Case  (3), also, satisfy \textit{simultaneity condition} as shown in  Fig. \ref{fig1new}(f). Figure \ref{fig7new}(a) shows the strength of the Yukawa couplings corresponding to the points satisfying  \textit{simultaneity condition}. In this  case $|y_1|$ varies in the range $(0.64-1.26)$. The hierarchy among Yukawa couplings exist as for without splitting case $(\delta M\simeq 0)$.  Since we have splitting between $N_1$ and $N_2$ at lower values of $r_1$ CDM relic density experimental constraint is not satisfied. It can be satisfied only when $r_1 \gtrsim 0.85$ (see Fig. \ref{fig7new}(b)). Corresponding to these points Yukawa coupling values are shown in Fig. \ref{fig7new}(c) as a function of $r_1$. The values of $m_0$ in this case lies in the range $(3300 - 5700)$ GeV. In fact like in NH case here we also have upper bound on mass of inert doublet $m_0$ around 5700 GeV as evident from Fig. \ref{fig7new}(d). Also from Fig. \ref{fig7new}(d) we can see that  MEG II experiment has enough sensitivity to test almost all region for $\mu \rightarrow e \gamma$ process and, if not found, then this case, for IH, in this model will be ruled out. The results and conclusions obtained for other LFV processes i.e., $\tau \rightarrow e \gamma$ and $\tau \rightarrow \mu \gamma$,  are not significantly different from the case without splitting. The values of splitting till which it can satisfy the  \textit{simultaneity condition} is from $(10\%- 15.5\%)$ approximately (see Fig. \ref{fig7new}(e)). The dark matter mass ranges from $(2900 - 5000)$ GeV and Higgs quartic coupling $\lambda_5$ from $(8\times 10^{-10}-2\times 10^{-9})$  as shown in Fig. \ref{fig7new}(f).
  
We will now study the neutrino phenomenology by considering the case (2) of possible coannihilation scenarios.
The benchmark point corresponding to $y_3=0.068+0.004 i$, $\alpha_1=49.512^o$ and $\alpha_2=26.856^o$  consistent with the \textit{simultaneity condition} is shown in fourth column of Table \ref{tab2}.

The similar trend as was observed for NH can, also, be seen here. The free parameters $\theta$ and $\phi$ exhibit similar behaviour, as that of NH, with two distinct regions, as evident from Fig. \ref{fig3}(a). The Jarlskog CP invariant, as depicted in Fig. \ref{fig3}(b), varies between $(-0.036$ to $0.036)$. The Dirac CP-violating phase $\delta$ varies between $(-90^o$ to $90^o)$.

In this case, the effective Majorana mass $|m_{ee}|$ is constrained to take values from $(0.015$ - $0.12)$ eV, and the sum of neutrino masses $\sum_{i} |m_i|$ takes values from $(0.1$ - $0.4)$ eV, as shown in Fig. \ref{fig3}(c). However, taking cosmological bound on sum of neutrino masses $\sum_{i} |m_i|$ into consideration, the \textit{simultaneity condition} predicts $|m_{ee}|\in (0.015-0.05)$ eV range which is well within the sensitivity limits of the $0\nu \beta \beta$ decay experiments. The non-observation of this process shall rule out IH predicted by the model. Furthermore, if the bound coming from cosmological observations on $\sum_{i} |m_i|$ becomes more stringent ($\backsimeq 0.1$ eV), the IH shall again be ruled out.

The Majorana CP invariants $I_1$ and $I_2$, along with their corresponding Majorana phases $\alpha_{21}$ and $\alpha_{31}$, are shown in Figs. \ref{fig3}(d) and \ref{fig3}(e), respectively. The range obtained for $I_1$ is $(-0.46$ to $0.46)$ and for $I_2$, it is $(-0.12$ to $0.12)$ approximately. The Majorana phases $\alpha_{21}$ and $\alpha_{31}$ vary in the entire range from $0^o$ to $360^o$. 

In Figs. \ref{fig3}(f) and \ref{fig3}(g), the correlation of Majorana CP phases with the effective Majorana mass $|m_{ee}|$ are shown. Figure \ref{fig3}(h), explains the behaviour of neutrino masses $|m_3|,|m_2|$, and $|m_1|$ with $\lambda_5$. The spread in their values increases as $\lambda_5$ increases, specifically when $\lambda_5>1.0\times 10^{-9}$. In this region, cosmological bound on sum of neutrino masses is, generally, satisfied. This is because  small values of the lightest mass eigenvalue $|m_3|$ prevent the sum of neutrino masses from taking higher values, even if the mass of two heavier mass eigenvalues in the spectrum (i.e., $|m_1|, |m_2|$) become close to 0.06 eV. Thus, cosmological bound is mostly satisfied for larger values of $\lambda_5$.

\begin{table}
    \centering
    \begin{tabular}{|c|c|c|c|c|} \hline 
    	S.No.&Parameters&NH&IH&NH \\
    		& & & &  (Extended Magic Symmtery) \\ \hline 
         1.&$\theta\, (^o)$&169.988&190.291& 10.173\\
         2.&$\phi\,(^o)$&208.659&282.328&244.172\\
         3.&$\lambda_2$&0.450&0.323&0.878\\
         4.&$\lambda_3$&0.063&0.798&0.143\\
         5.&$\lambda_4$&2.721$\times$ 10$^{-8}$&1.992$\times$ 10$^{-8}$&1.827$\times$ 10$^{-8}$\\
         6.&$\lambda_5$&8.059$\times$ 10$^{-10}$&1$\times$ 10$^{-9}$&1.315$\times$ 10$^{-9}$\\
         7.&$m_{\zeta}$&4.864$\times$ 10$^{3}$&4.212$\times$ 10$^{3}$&5.498$\times$ 10$^{3}$\\
         8.&$M_1$ (GeV)&4559.765&3812.513&4971.564\\
         9.&$M_3$ (GeV)&6832.374&5505.369&9380.505\\
         10.&$\Delta m^2_{21}\,(\text{eV}^2)$&7.98$\times$10$^{-5}$&7.09$\times$10$^{-5}$&7.03$\times$10$^{-5}$\\
         11.&$\Delta m^2_{31}\,(\text{eV}^2)$&2.49$\times$10$^{-3}$&2.52$\times$10$^{-3}$&2.56$\times$10$^{-3}$\\\hline
         12.&$\theta_{13}\, (^o)$&8.16&8.39&8.29\\
         13.& $\theta_{12}\, (^o)$&35.7&35.7&35.7\\
         14.&$\theta_{23}\, (^o)$&50.10&46.27&42.44\\
         15.&$\Omega h^2$&0.120&0.119&0.117\\
         16.&Br$(\mu\rightarrow e \gamma)$&3.23$\times$10$^{-14}$&1.39$\times$10$^{-13}$&3.09$\times$10$^{-14}$\\
         \hline
    \end{tabular}
    \caption{Benchmark points consistent with \textit{simultaneity condition} \textit{i.e.} satisfying neutrino oscillation data at $3\sigma$, experimental bounds on $\Omega h^2$ and Br($\mu\rightarrow e \gamma$). The point shown in last column, also, satisfy the extended magic symmetry.}
    \label{tab2}
\end{table}

 \begin{figure}[t]
   \centering
   \includegraphics[width=0.45\linewidth]{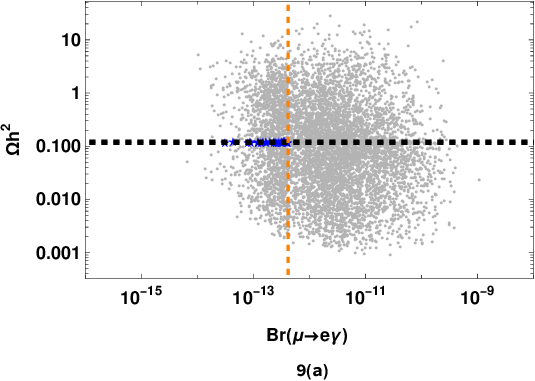}
 \begin{tabular}{cc}
  \includegraphics[width=0.45\linewidth]{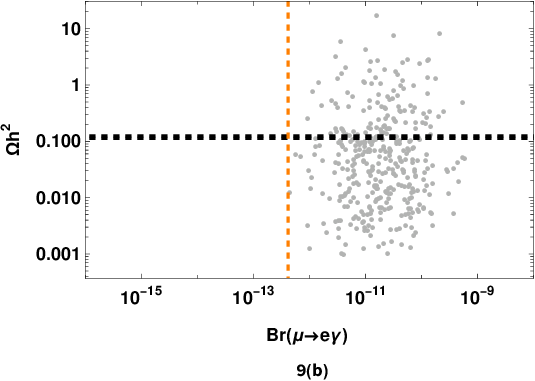}
&\includegraphics[width=0.45\linewidth]{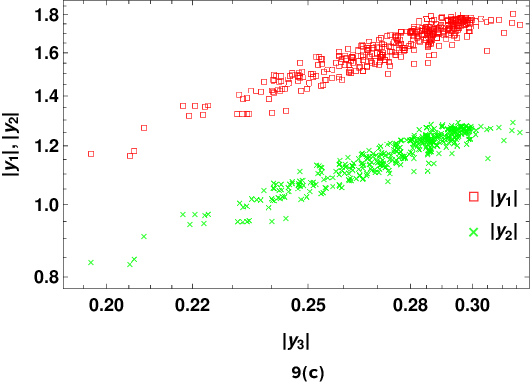}\end{tabular}
 \caption{ Case (2): For NH (first row)  and IH  (second row), all points shown satisfies neutrino oscillation data at $3\sigma$ with extended magic symmetry condition. The horizontal black  and vertical orange dashed lines  are experimental range of $\Omega h^2$ and experimental upper bound on Br($\mu\rightarrow e \gamma$) respectively. The points shown in blue color  (``\textcolor{blue}{$\filledstar$}") also satisfy the \textit{simultaneity condition}. Figure \ref{fig4}(c) shows strength of Yukawa couplings corresponding to all points in Fig. \ref{fig4}(b).  }
   \label{fig4}
 \end{figure}

 \begin{figure}[t]
	\centering
	\begin{tabular}{cc}
		\includegraphics[width=0.45\linewidth]{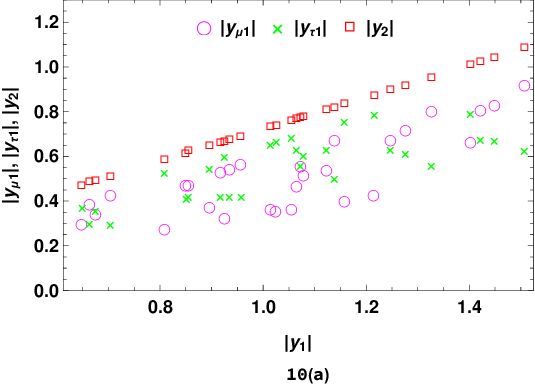}& \includegraphics[width=0.45\linewidth]{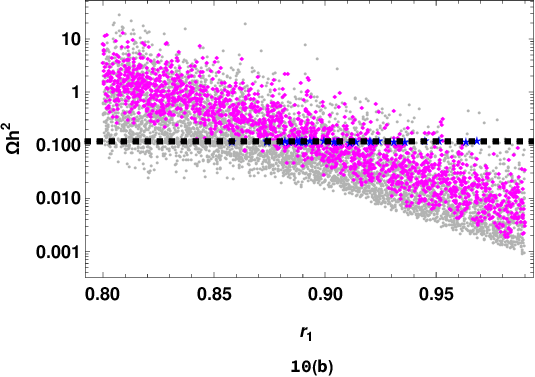}\\
		\includegraphics[width=0.45\linewidth]{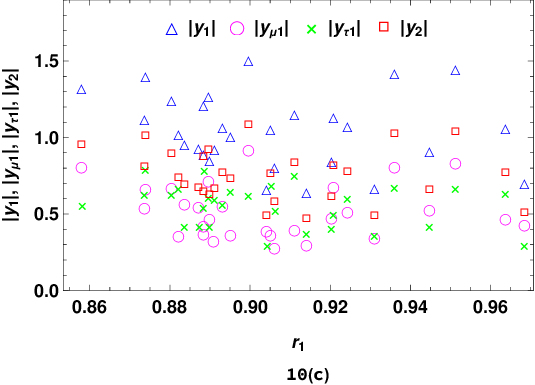}& \includegraphics[width=0.45\linewidth]{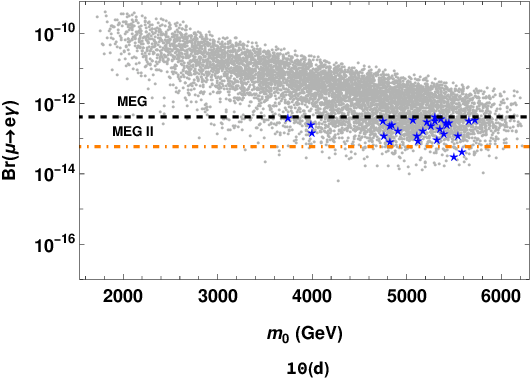}\\
		\includegraphics[width=0.45\linewidth]{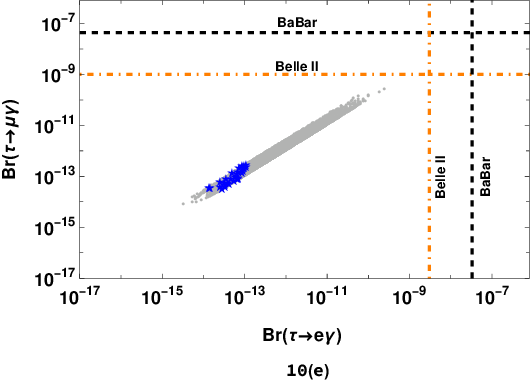}&\includegraphics[width=0.45\linewidth]{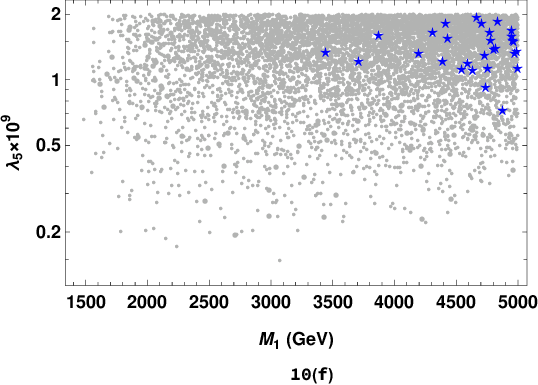}
	\end{tabular}
	\caption{Case (2) for NH: Figure \ref{fig10new}(a) and \ref{fig10new}(c) show the Yukawa coupling strength for points satisfying the \textit{simultaneity condition}. In Fig. \ref{fig10new}(b), diamond-shaped magenta points indicate points satisfying the experimental upper bound on Br($\mu \rightarrow e \gamma$), and black dashed horizontal lines represent the experimentally allowed 3$\sigma$ region for $\Omega h^2$. In Fig. \ref{fig10new}(d) and \ref{fig10new}(e), the black dashed and orange dotted-dashed lines show the sensitivities of the current and future  experiments, respectively.}
	\label{fig10new}
\end{figure}

\begin{figure}[t]
  \centering
\begin{tabular}{cc} 
	\includegraphics[width=0.45\linewidth]{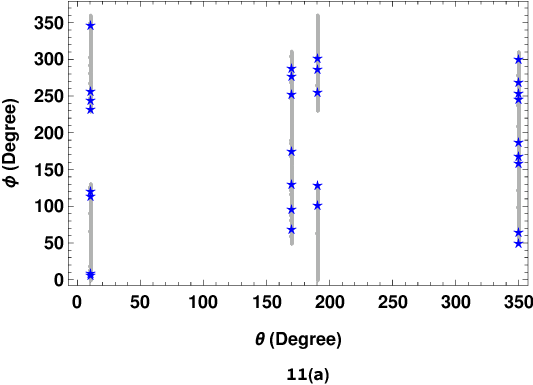}& \includegraphics[width=0.45\linewidth]{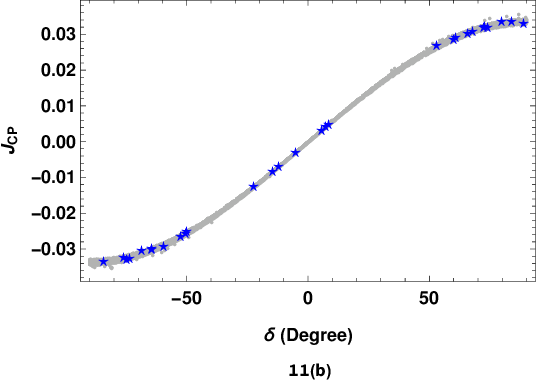}\\ 
	\includegraphics[width=0.45\linewidth]{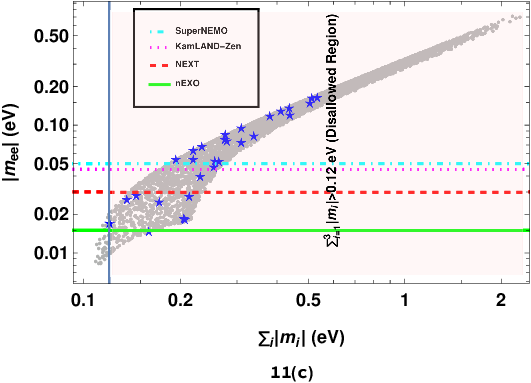}& \includegraphics[width=0.45\linewidth]{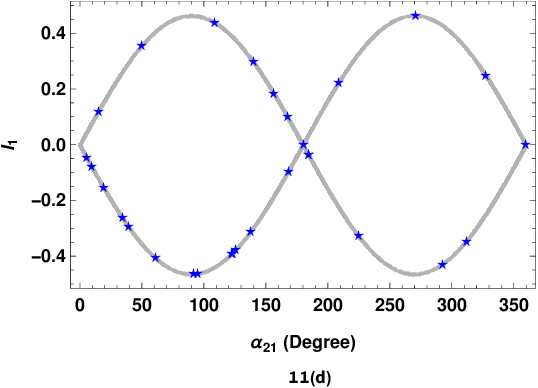}\\ 
	\includegraphics[width=0.45\linewidth]{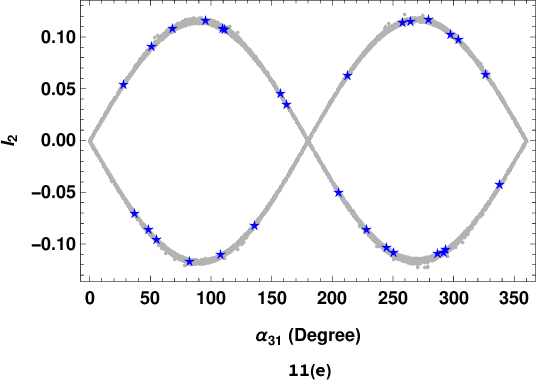}& \includegraphics[width=0.45\linewidth]{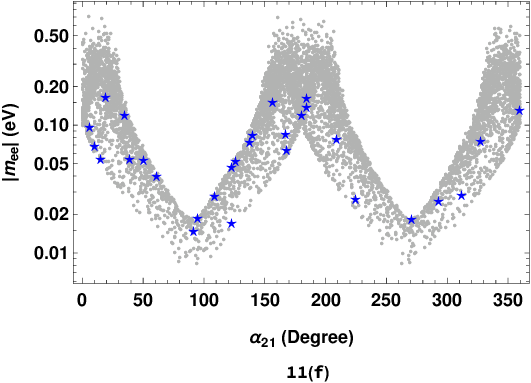}\\ 
	\includegraphics[width=0.45\linewidth]{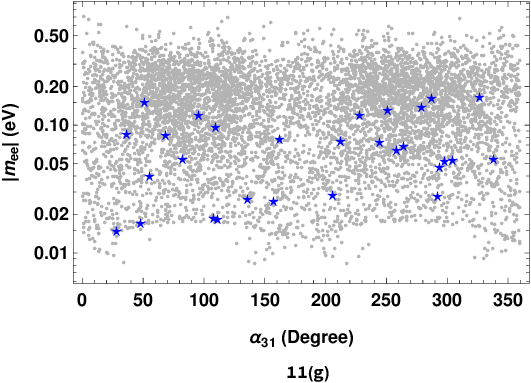}& \includegraphics[width=0.45\linewidth]{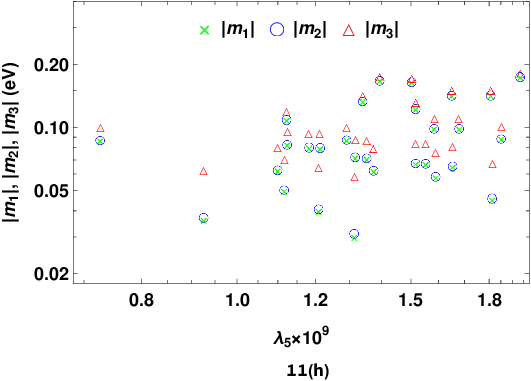}\\ \end{tabular}
  \caption{Case (2) for NH: The grey points show the parameter space that satisfies extended magic symmetry condition and neutrino oscillation data at $3\sigma$,  while the points shown in blue color  (``\textcolor{blue}{$\filledstar$}") satisfy the \textit{simultaneity condition} in addition to extended magic symmetry condition. In Fig. \ref{fig5}(h) all points satisfy \textit{simultaneity condition} along with  extended magic symmetry condition. }
  \label{fig5}
\end{figure}


\section{Extended Magic Symmetry}{\label{s6}}
Hitherto, we have adopted the trimaximal (TM$_2$) structure for the neutrino mixing matrix together with the constraints originating from the diagonalization condition of the neutrino mass matrix $m_{\nu}$. This adoption allows us to parameterize the Yukawa coupling matrix $y$ in terms of three independent Yukawa couplings: $y_1$, $y_2$, and $y_3$.
Utilizing this parameterized Yukawa coupling matrix $y$, we derive the neutrino mass matrix  $m_{\nu}$. Due to the TM$_2$ mixing scheme,  $m_{\nu}$ is a magic matrix with a magic sum equal to $m_2$ (see Eqn. (\ref{mv})). The TM$_2$ matrix diagonalizes  $m_{\nu}$, yielding the neutrino masses $m_1$, $m_2$, and $m_3$ which depends on Yukawa couplings $y_1$, $y_2$, and $y_3$ respectively (see Eqn. (\ref{m1m2m3_1})).  To be phenomenologically viable, these masses must satisfy the mass-squared differences, which further reduces the number of independent Yukawa coupling parameters to one, $y_1$ ($y_3$) for NH (IH).
Using the resulting matrix  $m_{\nu}$, we investigate neutrino phenomenology, relic density of DM and possible LFV in $\mu\rightarrow e\gamma$ process. 

\noindent As mentioned earlier, it is possible to assume certain structures or patterns in $m_{\nu}$, which can allow us to completely determine the remaining Yukawa coupling $y_1$ ($y_3$). Building on our previous work\cite{Singh:2022nmk}, wherein we have proposed an \textit{ansatze} for $m_\nu$ with an extended magic symmetry, we explore its impact on prediction of the model discussed in Section \ref{s5}.

\textit{Extended Magic Symmetry}: Under this extension of the TM$_2$ mixing scenario, (2,2) element of the neutrino mass matrix is equal to the magic sum \textit{i.e.} $m_2$. The symmetry origin of this mass relation has been presented in Ref. \cite{Singh:2022nmk} with $\Delta(54)$ discrete flavor symmetry\footnote{The Scotogenic origin of this ansatz is under investigation.}.

\noindent So, we have $(m_{\nu})_{22}$ element of the neutrino mass matrix (see Eqn. (\ref{mv}) ) equal to the magic sum $m_2$ which in the present model translates to the following form:
\begin{eqnarray}
        a_1^2 y_{1}^2 \Lambda_1+y_{2}^2 \Lambda_2+a_3^2 y_{3}^2 \Lambda_3 &=&c_2y_2^2 \Lambda_2,\nonumber \\
         a_1^2 y_{1}^2 \Lambda_1+y_{2}^2 \Lambda_2+a_3^2 y_{3}^2 \Lambda_3 -c_2 y_2^2 \Lambda_2&=&0.
\end{eqnarray}
For NH, using Eqns. (\ref{y21}) and (\ref{y31}), we can rewrite the above equation as
\begin{small}
  \begin{equation}\label{magic=mv22NH}
  \begin{split}
    a_1^2 y_{1}^2 \Lambda_1+\frac{1}{|c_2|}\sqrt{\Delta m^2_{21}+ |c_1  y_1^2 \Lambda_1 |^2}e^{ 2 i \alpha_2}+a_3^2\frac{1}{|c_3|}\sqrt{\Delta m^2_{31}+ |c_1  y_1^2 \Lambda_1 |^2}e^{2 i\alpha_3} -\sqrt{\Delta m^2_{21}+ |c_1  y_1^2 \Lambda_1 |^2}e^{ 2 i \alpha_2} =0.
  \end{split}
  \end{equation}
\end{small}
and, for IH, using Eqns. (\ref{y13}) and (\ref{y23}), we have
\begin{small}
	\begin{equation}\label{magic=mv22IH}
		\begin{split}
			a_1^2 \frac{1}{|c_1|}\sqrt{\Delta m^2_{31}+ |c_3  y_3^2 \Lambda_3 |^2}e^{2 i \alpha_1}&+\frac{1}{|c_2|}\sqrt{\Delta m^2_{21}+\Delta m^2_{31}+ |c_3  y_3^2 \Lambda_3 |^2} e^{2 i \alpha_2}+a_3^2 y^2_3 \Lambda_3\\
			&- \sqrt{\Delta m^2_{21}+\Delta m^2_{31}+ |c_3  y_3^2 \Lambda_3 |^2}e^{2 i \alpha_2}=0.
		\end{split}
	\end{equation}
\end{small} 

We can see from these equations that $y_1$ becomes a function of $\lambda_5$, $r_1$, and $m_0$ (through $\Lambda_1$), while $y_3$ becomes a function of $\lambda_5$, $r_3$, and $m_0$ (through $\Lambda_3$). Both depend on the free parameters $\theta$ (through $a_1$, $a_3$, $c_1$, and $c_3$) and $\phi$ (through $a_1$ and $a_3$). Also, dependence of $y_1$ ($y_3$) on phases $\alpha_2$ and $\alpha_3$ ( $\alpha_1$ and $\alpha_2$) is clearly visible. In this case, $y_1$ and $y_3$ are not free anymore; they have to satisfy constraints originating from extended magic symmetry.


\subsection{Numerical Analysis and Discussion}

In this section, we vary the neutrino mass-squared differences, $\theta$, $\phi$, and the parameters of the Scotogenic model within the same range as mentioned in the previous section. Yukawa couplings $y_1$ and $y_3$ are not free parameters anymore but are solutions to Eqns. (\ref{magic=mv22NH}) and (\ref{magic=mv22IH}), respectively. We solve these equations numerically considering those values of the Yukawa couplings as solutions for which $\left|\frac{(m_{\nu})_{22} - m_2}{m_2}\right| \leq 10^{-2}$ and $0 \leq |y_1|, |y_3| \leq 1.8$. Furthermore, we vary $\alpha_2$ and $\alpha_3$ in NH ($\alpha_1$ and $\alpha_2$ in IH), randomly in the range from $0$ to $2 \pi$, with uniform distribution.

\subsection{Normal Hierarchy}
In light of the discussion in the  Subsection  \ref{s5.1} case (1) is ruled out and we are left with cases (2) and (3) only which satisfy the \textit{\textit{simultaneity condition}}. We shall discuss only case (2), as the results obtained for case (3) are not significantly different, as evident from the analysis and results discussed in Subsection  \ref{s5.1}.\\
 In Fig. \ref{fig4}(a), correlation of CDM relic density  ($\Omega h^2$) and Br($\mu \rightarrow e \gamma$) is shown for all those points which satisfy  extended magic symmetry condition.  The 3$\sigma$ experimental range of CDM relic density is shown by black horizontal dashed line while orange vertical dashed line represents the experimental upper bound on Br($\mu \rightarrow e \gamma$).
As evident from this figure parameter space exist for which \textit{simultaneity condition} can be satisfied. The minimum possible value of Yukawa coupling $|y_1|$ is pushed to a higher value around 0.65 which in previous case (i.e., without extended magic symmetry) was  around 0.3. This is because of the extended magic symmetry condition which requires high value of Yukawa coupling $|y_1|$ (see Fig. \ref{fig10new}(a)). The values of $r_1$ after which \textit{simultaneity condition} can be satisfied remains same as previous case i.e., $r_1\gtrsim 0.86$ (see Fig. \ref{fig10new}(b)). Corresponding  strength of Yukawa couplings are shown in  Fig. \ref{fig10new}(c) as a function of $r_1$. While inert scalar mass varies from $(3700 -5700)$ GeV approximately. As in without extended magic symmetry case we have upper bound on $m_0$ with value around 5700 GeV (see Fig. \ref{fig10new}(d)). Now Br($\mu \rightarrow e \gamma$) is  restricted to take minimum value $\mathcal{O}(10^{-14})$ which lies just below the future sensitivity of the MEG II experiment (see Fig. \ref{fig10new}(d)). So, most part of Br($\mu \rightarrow e \gamma$) can be tested in MEG II experiment. The reason that Br($\mu \rightarrow e \gamma$) now cannot go below $\mathcal{O}(10^{-14})$  is that low values of Yukawa couplings are not available as we have already discussed. As far as Br($\tau \rightarrow e \gamma$) and Br($\tau \rightarrow \mu \gamma$) are considered the situation still remains the same with values  far below the experimental sensitivity of future experiments (see  Fig.  \ref{fig10new}(e)).  Also, the dark matter mass ranges from $(3400-5000)$ GeV and Higgs quartic coupling $\lambda_5$ has range from $(7\times 10^{-10}-2\times 10^{-9})$ as shown in Fig. \ref{fig10new}(f).


By considering this case we have studied neutrino phenomenology  and obtained the benchmark point corresponding to $y_1=-0.299-0.592 i$, $\alpha_2=185.525^o$ and $\alpha_3=111.380^o$ which satisfy the \textit{simultaneity condition} along with the extended magic condition, as shown in fifth column of Table \ref{tab2}. The results obtained are displayed in Fig. \ref{fig5}. Since we are imposing two conditions, namely the \textit{simultaneity condition} and the extended magic symmetry, it is expected that some of the regions or points allowed by the \textit{simultaneity condition} alone will be filtered out by the extended magic condition. Consequently, the number of points obtained is considerably fewer compared to when only one condition is imposed. As observed in correlation plots shown in Fig. \ref{fig5},  the trend is quite similar to NH without extended magic symmetry.

The values of the effective Majorana mass $|m_{ee}|$ and the sum of neutrino masses $\sum_{i} |m_i|$ are constrained by both conditions, as depicted in Fig. \ref{fig5}(c). The allowed points are now constrained within a very small region, as most of the part is disallowed by cosmological data for the sum of neutrino masses. These values of $|m_{ee}|$ is within reach of current and future $0\nu\beta\beta$ decay experiments. The non-observation of this decay process will refute case (2) for NH, in this model. Figs. \ref{fig5}(f) and \ref{fig5}(g) depict the correlation plots of $\alpha_{21}$ and $\alpha_{31}$ with $|m_{ee}|$, respectively. In Fig. \ref{fig5}(h), the plot showcases the relationship between the Higgs quartic coupling $\lambda_5$ and three neutrino masses, where each point, represented by different colors, satisfies the \textit{simultaneity condition} and the extended magic condition. It is to be noted that extended magic symmetry scenario prefers higher values of $\lambda_5$ as compared to the, only-TM$_2$ scenario discussed in Subsection \ref{s5.1}.

\subsection{Inverted Hierarchy}

In Fig. \ref{fig4}(b), the points  in $\Omega h^2$ and Br($\mu \rightarrow e \gamma$) plane are those for which extended magic symmetry condition is satisfied.  As depicted in Fig. \ref{fig4}(b), there are no points below the experimental upper bound for Br($\mu \rightarrow e \gamma$). This observation is remarkable as it indicates that if both these conditions (simultaneity and extended magic conditions) need to be fulfilled simultaneously, then IH is ruled out. This happens because, in order to satisfy the extended magic symmetry condition Yukawa couplings are required to have higher values. In IH case,  $y_3$ is the indepndent Yukawa coupling and if we see the relations (Eqns. (\ref{y13}) and (\ref{y23})) then we can see that $y_3$ cannot take  much higher values otherwise Yukawa couplings $y_1$ and $y_2$ will violate the perturbativity limit. So, its values are controlled by the extended magic symmetry from below, as it prevents  $y_3$  from taking lower values, and by the perturbativity limit from above, as it stops  $y_3$ from taking higher values  (see Fig. \ref{fig4}(c)). 
\section{Conclusions}{\label{s7}}

In conclusions, we have studied the Scotogenic model where we have used trimaximal mixing matrix $U_{\text{TM}_2}$ as the neutrino mixing matrix, and using the diagonalization condition for the flavor neutrino mass matrix $m_{\nu}$, we have parameterized the Yukawa coupling matrix $y$. We have investigated the allowed parameter space of the model satisfying experimental bounds on relic density of CDM and possible LFV alongside neutrino oscillation data, which we called the \textit{simultaneity condition}, in the framework of trimaximal mixing paradigm. In particular, we have studied different cases by assuming three coannihilation scenarios in two settings: (i) wherein $m_\nu$ is magic symmetric, and (ii)  $m_\nu$ has extended magic symmetry \textit{i.e.}, $(m_{\nu})_{22}$ element of $m_\nu$ is, also, equal to the magic sum $m_2$.

In the first setting, we find that case (1), independent of neutrino mass hierarchy cannot satisfy the \textit{simultaneity condition} and hence ruled out. On the other hand the coannihilation scenarios case (2) and  (3) both are found to be consistent  with the  \textit{simultaneity condition}  in both normal and inverted hierarchies of neutrino masses. The combined analysis of cases (2) and (3) reveals  that, in both NH and IH, splitting between masses of $N_1$ and $N_2$  can be up to $\approx 15\%$ for the model to obey  \textit{simultaneity condition}. The value of mass ratio $r_1$ need to be greater than $0.85$. There is an upper bound on the mass of inert scalar $m_0$ around 5700 GeV. The allowed parameter space lies beyond the sensitivities of future experiments which can probe Br($\tau \rightarrow \mu \gamma$) and Br($\tau \rightarrow e \gamma$) parameter space.  However, as far as Br($\mu \rightarrow e \gamma$) is concerned, in NH, it can go as low as $\mathcal{O}(10^{-15})$, whereas in IH, it can barely go beyond the sensitivity of future experiment MEG II. So, MEG II can  probe the IH case and if not found then case (3) can be ruled out in IH, in this model. The dark matter mass lies in $(2500 -5000)$  GeV range. By considering case (2) to study neutrino phenomenology we have found out that both normal as well as inverted hierarchies of neutrino mass are allowed by the \textit{simultaneity condition}. For the free parameters $\theta$ and $\phi$ there exist two distinct regions of parameter space under simultaneity constraint \textit{viz.,} (i) for $\theta\backsimeq 10^o \text{ or } 190^o$: $\phi\in (0^o-130^o)\oplus (230^o-360^o)$ (ii) for $\theta\backsimeq 170^o \text{ or } 350^o$: $\phi\in (50^o-310^o)$ in both hierarchies. Also in both hierarchies, the Dirac phase $\delta$ and Majorana phases $\alpha_{21}$, $\alpha_{31}$ span the entire allowed range, with the Jarlskog invariant $J_{CP}$ varying in the range $(-0.036$ to $0.036)$ and the Majorana CP invariants $I_1$ and $I_2$ varying in the range $(-0.46$ to $0.46)$ and $(-0.12$ to $0.12)$, respectively. The effective Majorana mass $|m_{ee}|$ takes the range $(0.002$ to $0.15) \, \text{eV}$ for NH and $(0.015$ to $0.12) \, \text{eV}$ for IH.  However, a significant region of $|m_{ee}|$ is excluded by the cosmological bound on the sum of neutrino masses.  Considering the cosmological bound on sum of neutrino masses $\sum_{i} |m_i|$, the \textit{simultaneity condition} predicts $|m_{ee}|\in (0.002-0.03)$ eV range for NH and  $|m_{ee}|\in (0.015-0.05)$ eV range for IH.  The predicted range of  $|m_{ee}|$ can be probed at $0\nu \beta \beta$ decay experiments which is well within the sensitivity limits of the $0\nu \beta \beta$ decay experiments for IH. If this process is not observed, it will rule out the IH predicted by the model. Furthermore, if the upper limit derived from cosmological observations on the sum of neutrino masses ($\sum_{i} |m_i|$) becomes more restrictive ($\backsimeq 0.1$ eV), the inverted hierarchy will again be excluded.

In the second setting, NH is allowed for case (2) and case (3) while IH is disallowed in all cases of coannihilations considered in the present work. We have studied case (2) in detail for NH, where we have found out that due to extended magic symmetry condition parameter space for Br($\mu \rightarrow e \gamma$) is constrained, with its lowest value nearly below the experimental sensitivity of MEG II experiment. So, NH with extended magic symmetry can be tested in this experiment. The upper bound on mass inert scalar remains the same as without extended magic symmetry setting. The dark matter mass now ranges from $(3400-5000)$ GeV, approximately. The neutrino phenomenology is also studied for this case, and the obtained parameter space resembles with that of first setting. However, the range of $|m_{ee}|$ obtained in second setting, consistent with the cosmological bound on the sum of neutrino masses, is highly constrained. The more stringent bounds coming from cosmological data on the sum of neutrino masses in the future can test the viability of NH with extended magic symmetry. 

For NH, the obtained values of neutrino masses for the second setting take higher values which corresponds to the higher value of quartic coupling, $\lambda_5$. In turn, this pushes the sum of neutrino masses to go up and hence in this setting  we have more tight constraint from cosmological data on the sum of neutrino masses.

In summary, our investigation sheds light on the interplay between the Scotogenic model and TM$_2$ mixing leading to the reduction in the number of free parameters and stringent constraints on the allowed parameter space. The model is predicting NH of neutrino masses with both CP conserving and CP violating solutions along with exclusion of the inverted hierarchy in presence of the extended magic symmetry.

\noindent\textbf{\Large{Acknowledgments}}
 \vspace{.3cm}\\
Tapender acknowledges the financial support provided by Central University of Himachal Pradesh. The authors, also, acknowledge Department of Physics and Astronomical Science for providing necessary facility to carry out this work.

\end{document}